\documentclass[screen]{acmart}

\AtBeginDocument{%
  \providecommand\BibTeX{{%
    \normalfont B\kern-0.5em{\scshape i\kern-0.25em b}\kern-0.8em\TeX}}}

\usepackage[T1]{fontenc}
\usepackage[american]{babel}

\usepackage{xcolor}
\usepackage{graphicx}

\usepackage[printonlyused]{acronym}
\usepackage{environ}
\usepackage{amsmath}
\usepackage{amsfonts}
\usepackage{tikz}
\usetikzlibrary{fit,positioning,shapes,calc}
\usepackage{booktabs}
\usepackage{subcaption}
\usepackage{verbatim}
\usepackage{fancyvrb}
\usepackage{hyperref}
\usepackage{xspace}
\usepackage{proof}
\usepackage{xfrac}
\usepackage{rotating}
\usepackage{multirow}
\usepackage{listings}
\usepackage{hhline}
\usepackage[inline]{enumitem}
\usepackage{arydshln}
\usepackage{balance}

\usepackage[ruled]{algorithm2e}
\definecolor{maincolor}{HTML}{368869}
\definecolor{secondcolor}{HTML}{A9769C}
\colorlet{maincolor@light}{maincolor!19!white}
\colorlet{maincolor@dark}{maincolor!80!black}
\colorlet{secondcolor@light}{secondcolor!19!white}
\definecolor{secondcolor@dark}{HTML}{8a3073}

\newcommand{\rposts}{\texttt{recent-posts}\xspace}
\newcommand{\trposts}{\texttt{top-rated-posts}\xspace}
\newcommand{\textbook}{\texttt{textbook}\xspace}
\newcommand{\spider}{\texttt{spider}\xspace}
\newcommand{\kaggle}{\texttt{kaggle}\xspace}

\newcommand{\patsql}{\textsc{PatSQL}\xspace}
\newcommand{\squares}{\textsc{Squares}\xspace}
\newcommand{\scythe}{\textsc{Scythe}\xspace}
\newcommand{\cubes}{\textsc{Cubes}\xspace}
\newcommand{\cubesseq}{\mbox{\textsc{Cubes-Seq}}\xspace}

\newcommand{\cubesdcB}{\mbox{\textsc{Cubes-DC}}\xspace}
\newcommand{\cubesdc}[1]{\mbox{\textsc{Cubes-DC#1}}\xspace}

\newcommand{\pluseq}{\mathrel{+}=}

\acrodef{sygus}[SyGuS]{Syntax-Guided Synthesis}
\acrodef{SMT}{Satisfiability Modulo Theories}
\acrodef{SAT}{Propositional Satisfiability}
\acrodef{CEGIS}{Counter-Example Guided Inductive Synthesis}
\acrodef{PBE}{Programming by Example}
\acrodef{CSS}{Cascading Style Sheets}
\acrodef{EUF}{Equality with Uninterpreted Functions}
\acrodef{AST}{Abstract Syntax Tree}
\acrodef{CNF}{Conjunctive Normal Form}
\acrodef{DSL}{Domain Specific Language}
\acrodef{QRE}{Query Reverse Engineering}
\acrodef{LIA}{Linear Integer Arithmetic}
\acrodef{DFS}{Depth-first Search}
\acrodef{CDCL}{Conflict-driven Clause Learning}
\acrodef{VBS}{Virtual Best Solver}
\acrodef{LCDP}{Low-Code Development Platform}
\acrodef{PS}{Program Synthesis}
\acrodef{CFG}{Context-free Grammar}
\acrodef{SQL}{Structured Query Language}
\acrodef{CSV}{Comma-Separated Values}
\acrodef{NLP}{Natural Language Processing}
\acrodef{QF FD}{Quantifier-Free Finite Domain}
\acrodef{NL}{Natural Language}
\acrodef{KDE}{Kernel Density Estimation}

\NewEnviron{rcfg}{
    \newcommand{\produces}{&&\rightarrow}
    \newcommand{\alignprod}{& && \phantom{\rightarrow}}
    \newcommand{\term}[1]{\texttt{##1}}
    
    \newcommand{\prodtwo}[3]{\term{##1(}##2\term{,}\,##3\term{)}}
    \newcommand{\prodthree}[4]{\term{##1(}##2\term{,}\,##3\term{,}\,##4\term{)}}
    \newcommand{\prodfour}[5]{\term{##1(}##2\term{,}\,##3\term{,}\,##4\term{,}\,##5\term{)}}
    \newcommand{\ror}{\enspace|\enspace}
    \begin{alignat*}{2}
        \BODY
    \end{alignat*}
}

\tikzstyle{tex1} = [minimum width=3cm, minimum height=1cm, text centered]
\tikzstyle{tef1} = [minimum width=1.4cm, minimum height=.5ex]
\tikzstyle{tsml} = [minimum width=3cm, text centered, font=\footnotesize]
\tikzstyle{box1} = [rectangle, rounded corners, inner sep=.5em, text centered, draw=maincolor, fill=maincolor@light, thick]
\tikzstyle{box2} = [rectangle, inner sep=.5em, text centered, draw=secondcolor, fill=secondcolor@light, thick]
\tikzstyle{box1d} = [rectangle, rounded corners, inner sep=.5em, text centered, draw=maincolor, fill=maincolor@light, thick, dashed]

\tikzstyle{ktree} = [level distance=4.5em,
level 1/.style={sibling distance=12em},
level 2/.style={sibling distance=4em},
every node/.style={draw,circle,minimum width=2.75em},
every label/.style={draw=none,rectangle,text height=1.5ex}]

\newcommand*{\textcite}{\citet}

\makeatletter
\tikzset{%
     remember picture with id/.style={%
       remember picture,
       overlay,
       save picture id=#1,
     },
     save picture id/.code={%
       \edef\pgf@temp{#1}%
       \immediate\write\pgfutil@auxout{%
         \noexpand\savepointas{\pgf@temp}{\pgfpictureid}}%
     },
     if picture id/.code args={#1#2#3}{%
       \@ifundefined{save@pt@#1}{%
         \pgfkeysalso{#3}%
       }{
         \pgfkeysalso{#2}%
       }
     }
   }

   \def\savepointas#1#2{%
  \expandafter\gdef\csname save@pt@#1\endcsname{#2}%
}

\def\tmk@labeldef#1,#2\@nil{%
  \def\tmk@label{#1}%
  \def\tmk@def{#2}%
}

\tikzdeclarecoordinatesystem{pic}{%
  \pgfutil@in@,{#1}%
  \ifpgfutil@in@%
    \tmk@labeldef#1\@nil
  \else
    \tmk@labeldef#1,(0pt,0pt)\@nil
  \fi
  \@ifundefined{save@pt@\tmk@label}{%
    \tikz@scan@one@point\pgfutil@firstofone\tmk@def
  }{%
  \pgfsys@getposition{\csname save@pt@\tmk@label\endcsname}\save@orig@pic%
  \pgfsys@getposition{\pgfpictureid}\save@this@pic%
  \pgf@process{\pgfpointorigin\save@this@pic}%
  \pgf@xa=\pgf@x
  \pgf@ya=\pgf@y
  \pgf@process{\pgfpointorigin\save@orig@pic}%
  \advance\pgf@x by -\pgf@xa
  \advance\pgf@y by -\pgf@ya
  }%
}

\newcommand*{\@DrawBoxHeightSep}{0.3em}%
\newcommand*{\@DrawBoxDepthSep}{0.2em}%
\newcommand{\@DrawBox}[6][red]{
    \tikz[overlay,remember picture,baseline]
    \draw[thick,draw,rounded corners=.1em,#1]
      ($(pic cs:#4)+(-0.2em,#2+\@DrawBoxHeightSep)$) rectangle
      ($(pic cs:#5)+(0.2em,-#3-+\@DrawBoxDepthSep)$);
      \tikz[overlay,remember picture,baseline]\node[anchor=base] at ($(pic cs:#4)!0.5!(pic cs:#5)$) {#6};
}

\newcommand\tikzmark[2][]{%
\tikz[remember picture with id=#2] #1;}

\newcounter{image}
\newdimen\@myBoxHeight%
\newdimen\@myBoxDepth%
\newcommand{\tikzhighlight}[2][red]{%
    \settoheight{\@myBoxHeight}{#2}
    \settodepth{\@myBoxDepth}{#2}
    \tikzmark{l\theimage}#2\tikzmark{r\theimage}\@DrawBox[#1]{\@myBoxHeight}{\@myBoxDepth}{l\theimage}{r\theimage}{#2}
    \stepcounter{image}
}
\makeatother

\lstdefinestyle{mystyle}{
    basicstyle=\ttfamily,
}

\newlength\mylen
\newcommand\myinput[1]{%
  \settowidth\mylen{\KwIn{}}%
  \setlength\hangindent{\mylen}%
  \hspace*{\mylen}#1\\}
  
  \newcommand{\negphantom}[1]{\settowidth{\dimen0}{#1}\hspace*{-\dimen0}}

\hypersetup{draft}

\begin{document}
\title{\cubes: A Parallel Synthesizer for SQL Using Examples}

\author{Ricardo Brancas}
\affiliation{%
\institution{INESC-ID / Instituto Superior Técnico, Universidade de Lisboa}
\state{Portugal}
}
\email{ricardo.brancas@tecnico.ulisboa.pt}

\author{Miguel Terra-Neves}
\affiliation{%
\institution{OutSystems}
\state{Portugal}
}
\email{miguel.neves@outsystems.pt}

\author{Miguel Ventura}
\affiliation{%
\institution{OutSystems}
\state{Portugal}
}
\email{miguel.ventura@outsystems.pt}

\author{Vasco Manquinho}
\affiliation{%
\institution{INESC-ID / Instituto Superior Técnico, Universidade de Lisboa}
\state{Portugal}
}
\email{vasco.manquinho@tecnico.ulisboa.pt}

\author{Ruben Martins}
\affiliation{%
\institution{Carnegie Mellon University}
\state{USA}
}
\email{rubenm@cs.cmu.edu}

\renewcommand{\shortauthors}{R. Brancas, M. Terra-Neves, M. Ventura, V. Manquinho, R. Martins}

\begin{abstract}
In recent years, more people have seen their work depend on data manipulation tasks. However, many of these users do not have the background in programming required to write complex programs, particularly SQL queries. One way of helping these users is automatically synthesizing the SQL query given a small set of examples. Several program synthesizers for SQL have been recently proposed, but they do not leverage multicore architectures.

This paper proposes \cubes, a parallel program synthesizer for the domain of \acs*{SQL} queries using input-output examples. Since input-output examples are an under-specification of the desired SQL query, sometimes, the synthesized query does not match the user's intent. \cubes incorporates a new disambiguation procedure based on fuzzing techniques that interacts with the user and increases the confidence that the returned query matches the user intent.

We perform an extensive evaluation on around 4000 SQL queries from different domains. Experimental results show that our sequential version can solve more instances than other state-of-the-art SQL synthesizers. Moreover, the parallel approach can scale up to 16 processes with super-linear speedups for many hard instances. Our disambiguation approach is critical to achieving an accuracy of around 60\%, significantly larger than other SQL synthesizers.
\end{abstract}

\maketitle

\section{Introduction}
\label{sec:intro}

In the age of digital transformation, many people are being reassigned to tasks that require familiarity with programming or database usage. However, many users lack the technical skills to build queries in a language such as \ac{SQL}.
As a result, several new systems have been proposed for automatically generating SQL queries for relational databases~\cite{scythe,orvalhoVLDB20,ratsql,sqlizer}.
The goal of \emph{query synthesis} is to automatically generate an SQL query that
corresponds to the user's intent. For instance, the user can specify their intent using natural language~\cite{ratsql,sqlizer} or examples~\cite{sql-output,scythe,orvalhoVLDB20,patsql}. In this paper, we target query synthesis using examples, where each example consists of a database and an output table that results from querying the database. This problem of synthesizing SQL queries from input-output examples is known as Query Reverse Engineering.

\autoref{fig:example} illustrates an input-output example with two input 
tables (Courses and Grades) and an output table. The output table corresponds to 
counting the number of grades in each course. In this example, the goal is to 
synthesize the following SQL query:

\begin{figure}[t]
    \centering
    \begin{subfigure}[c]{.48\linewidth}
        \centering
        \begin{tabular}{ c c c }
        \toprule
        CourseID & StudentID & Grade \\ \midrule
        10 & 36933 & A \\
        11 & 36933 & B \\
        12 & 36933 & A \\
        10 & 37362 & A \\
        12 & 37362 & C \\
        11 & 37453 & A \\
        10 & 37510 & B \\
        12 & 37510 & A \\
        10 & 37955 & A \\ \bottomrule
        \end{tabular}
        \caption{The \texttt{Grades} table.} \label{fig:example-b}
    \end{subfigure}%
    \begin{minipage}{.5\linewidth}
        \begin{subfigure}[c]{\linewidth}
            \centering
            \begin{tabular}{ c c }
            \toprule
            CourseID & CourseName \\ \midrule
            10 & Programming \\
            11 & Algorithms \\
            12 & Databases \\ \bottomrule
            \end{tabular}
            \caption{The \texttt{Courses} table.} \label{fig:example-a}
        \end{subfigure}
        \par\medskip
        \begin{subfigure}[c]{\linewidth}
            \centering
            \vspace{1ex}
            \begin{tabular}{ c c }
            \toprule
            CourseName & GradeCount \\ \midrule
            Programming & 4 \\
            Algorithms & 2 \\
            Databases & 3 \\ \bottomrule
            \end{tabular}
            \caption{The output table.} \label{fig:example-c}
        \end{subfigure}
    \end{minipage}
    
    \caption{Two input tables: \texttt{Courses} and \texttt{Grades}. Output table: number of grades per course.}  
    \label{fig:example}
\end{figure}

\begin{lstlisting}{sql}
SELECT CourseName, count(*) AS 'GradeCount' 
FROM Grades
    NATURAL JOIN Courses
GROUP BY CourseName
\end{lstlisting}

Observe that, for a person with limited database training, in many situations, it is
easier to define one or more examples than to learn how to write the desired SQL query. 

Even though query synthesis tools using examples~\cite{sql-output,scythe,orvalhoVLDB20,patsql} have seen a remarkable improvement in recent years, they still suffer from scalability problems with respect to the size of the input tables and the complexity of the synthesized queries. 
Nowadays, multicore processors have become the predominant architecture for common laptops and servers. However, none of the previous query synthesis tools take advantage of the parallelism available in these architectures. In this work, we present \cubes, the \emph{first parallel synthesizer} for \ac{SQL} queries. \cubes is built on top of an open-source sequential query synthesizer~\cite{orvalhoVLDB20}, which we further improved by extending the language of queries supported by \cubes and by adding pruning techniques that can prevent incorrect programs from being enumerated. To take advantage of parallel architectures, we extend \cubes by using \emph{divide-and-conquer}. In this approach, each process searches a smaller sub-problem until it either finds a solution or exhausts that subspace and chooses another sub-problem to solve. We present a novel approach to create sub-problems based on considering different subsets of the domain-specific language for each process. 

To evaluate our tool, we collected benchmarks from prior work~\cite{scythe, sql-output, patsql, orvalhoVLDB20}. Also, we created a new dataset by extending existing query synthesis problems using natural language~\cite{spider} to use examples instead. In the end, we collected around 4000 instances that will be publicly available and can be used by other researchers when evaluating query synthesis tools using examples. 

We perform an exhaustive comparison between \cubes and state-of-the-art SQL synthesizers based on examples~\cite{scythe,orvalhoVLDB20,patsql}. Our evaluation shows that current SQL synthesizers can synthesize many SQL queries that satisfy the examples but do not match the user intent. We observe that \emph{all} state-of-the-art SQL synthesizers return fewer than 50\% of queries that match the user intent, i.e., even though they satisfy the example given by the user they do not match the query that the user had in mind. \cubes addresses this challenge by using parallelism to find multiple solutions and interact with the user to \emph{disambiguate} the query that matches the user intent. To disambiguate the queries, we use fuzzing to produce new examples that result in a different output for the possible synthesized queries. We select one of these examples and ask the user if the output is correct for these new input tables. If the user responds affirmatively, we can discard all queries that do not match this new output. Otherwise, we can discard the queries that match the new output. We repeat this process until we are confident that we found the query the user intended.

To summarize, this paper makes the following key contributions:

\begin{itemize}
    \item We improve sequential query synthesizers with new pruning techniques (\autoref{sec:sequential}).
    \item We propose a parallel approach for query synthesis based on divide-and-conquer techniques (\autoref{sec:parallel}).
    \item We propose a disambiguation procedure that uses fuzzing to disambiguate a set of queries that satisfies the initial example (\autoref{sec:disambiguation}).
    \item We created the largest dataset for query synthesis using examples with around 4000 instances (\autoref{sec:evaluation}).
    \item We implement the proposed ideas in a tool called \cubes and perform an extensive evaluation. The parallel version of \cubes with 16 processes outperforms the sequential version by solving 9\% more instances and having a median speedup of around 15$\times$ on hard instances solved by both approaches (\autoref{sec:evaluation}).
    \item We present the first study that analyses the accuracy of the queries returned by SQL synthesizers and show that more than 55\% of the queries do not match the user intent. Our disambiguation procedure can improve the accuracy of \cubes to 60\% and significantly outperform other example-based SQL synthesizers (\autoref{sec:evaluation}).
\end{itemize}

\section{Sequential Synthesis}
\label{sec:sequential}

This section describes the sequential synthesis procedure used in \cubes. Our framework builds upon \squares~\cite{orvalhoVLDB20}, a sequential query synthesizer that is modular and open-source which makes it easier to modify and extend to a parallel setting. First, we describe the general architecture of the sequential synthesizer. Next, we extend the sequential synthesizer in \cubes with a larger language and with a new pruning strategy. Finally, we compare the features and requirements of \cubes with those of other SQL synthesizers.

\subsection{Background Concepts}
We start by introducing background concepts on SQL program synthesis that will be used to describe \cubes and other related synthesizers.

\paragraph{\ac{DSL}} In general, synthesizers cannot support the complete program space of a full-featured programming language due to tractability problems of the search space. As such, synthesizers adopt a subset of the target language that balances expressiveness with synthesis performance. 
These subsets are called \acfp{DSL}. Although the \ac{DSL} of a program synthesizer may be represented internally by a different intermediate representation, the output of the synthesizer is always an SQL query that the user can execute.

\paragraph{Specification} 
A \emph{specification} for an SQL synthesizer using examples consists of a \emph{single} input-output example. The input corresponds to a database, and the output to the expected result of running the intended query on that database. For usability purposes, it is common for the user to provide a small database with the same schema as the original database but with fewer contents. The synthesized SQL query is guaranteed to satisfy the user specification on the small database while being generic to be run on larger databases and still preserve the user intent.

\paragraph{Ground truth} A \emph{ground truth} SQL query for a given instance is a query that is considered a correct solution for that instance. It is a query that returns the output expected by the user. 

\paragraph{Sketch}
A \emph{sketch} is a partial query with some missing parts. These parts are called holes and are represented by \(\square\). Examples of sketches: \texttt{SELECT * FROM~\(\square\)}, where \(\square\) corresponds to a missing table; \texttt{SELECT~\(\square\)~FROM t1 GROUP BY~\(\square\)}, where both \(\square\) correspond to missing columns.

\paragraph{\acf{SMT}} \acs{SMT} is a generalization of Boolean Satisfiability (SAT) that allows the use of high-level concepts such as integers, real numbers, and lists. While SAT only allows propositional logic formulations, SMT formulas allow fragments of first-order logic. A solution for an \ac{SMT} formula, also known as a model of the formula, is an assignment to the variables such that the formula evaluates to True.

\paragraph{Enumeration-based Synthesis}
Enumeration-based program synthesizers iterate through the program space (as defined by the synthesizer's \ac{DSL}) until they find a query that satisfies the user's specification. Due to the program space being usually infinite, most synthesizers enumerate queries in an increasing number of DSL operators. This satisfies the Occam razor principle. The main challenges in enumeration-based synthesis are enumerating programs in a favorable order and reusing information from previously attempted queries to speed up the search.

\paragraph{Sketch-based Synthesis}
The search process in a sketch-based program synthesizer usually starts with the empty sketch, \(\square\). Then, the synthesizer iteratively attempts to replace holes in the sketch with new operators or arguments until a complete query that satisfies the specification is found. Due to the nature of the process, it is usually required to use techniques such as backtracking or beam search. Several challenges arise in sketch-based enumeration, such as choosing the next hole to be filled, how to fill it, and developing efficient ways to backtrack and search.

\subsection{Synthesizer Architecture}

In programming-by-example (PBE) synthesizers, the user provides
an input-output example as specification for the query to be synthesized.
Besides the input and output tables in the example, the user can specify
optional elements to the synthesizer, such as a list of possible aggregation
functions (e.g. \texttt{count}, \texttt{sum}, \texttt{avg}).
The synthesizer uses a \ac{DSL} to specify the space of possible programs.
Using an intermediate representation in R\footnote{Table manipulation operations are implemented using \texttt{dplyr}, 
from \texttt{tidyverse}: \url{https://www.tidyverse.org/}.} instead of SQL, allows for a more compact representation of complex SQL operators.

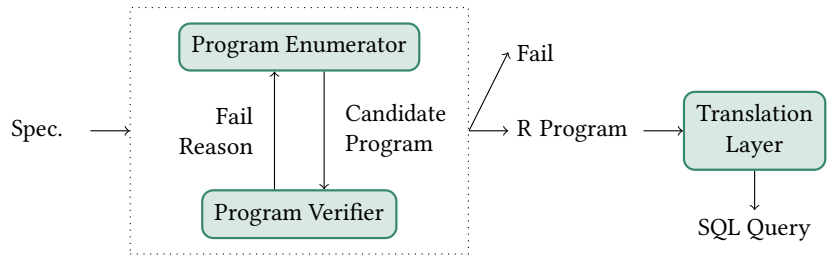
\begin{figure}[t]
    \resizebox{.75\linewidth}{!}{
\centering
\begin{tikzpicture}[node distance=1.5cm]

  \node[box1] (enum) [] {Program Enumerator};
  \node[box1] (dedu) [below = 1.5cm of enum] {Program Verifier};

  \draw[->]   ([xshift=1em]enum.south) -- ([xshift=1em]dedu.north) node[midway, right, xshift=.5em, align=left, text width=4em] (l1) {Candidate\\ Program};
  \draw[->]   ([xshift=-1em]dedu.north) -- ([xshift=-1em]enum.south) node[midway, left, xshift=-.5em, align=right, text width=4em] (l2) {Fail\\Reason};

  \node[draw,dotted,fit=(enum) (dedu) (l1) (l2), inner sep=.6em] (synt) {};

  \node[tef1] (spec) [left  = .5cm of synt, text width = 0.6cm, align=right] {Spec.};
  \node[tef1] (prog) [right = .5cm of synt, yshift=0cm, anchor=west, text width=1.5cm, align=left] {R Program};
  \node[tef1] (fail) [right = .5cm of synt, yshift=1cm, anchor=west, text width=1.5cm, align=left] {Fail};

  \draw[->]   (spec)  -- (synt);
  \draw[->]   (synt.east) -- (prog.west);
  \draw[->]   (synt.east) -- (fail.west);

  \node[box1] (trans) [right = .5cm of prog,align=center] {Translation\\Layer};
  \node[tef1] (query) [below = .5cm of trans] {SQL Query};

  \draw[->]   (prog)  -- (trans);
  \draw[->]   (trans)  -- (query);

\end{tikzpicture}}
    \caption{Diagram of sequential enumeration-based synthesizer architecture.} 
    \label{fig:squares}
\end{figure}

\autoref{fig:squares} presents the typical sequential synthesizer architecture 
of enumeration-based program synthesizers~\cite{orvalhoVLDB20}.
The synthesizer receives a specification from the user and performs a program
enumeration process (Program Enumerator). For each enumerated candidate program, it 
is verified if it satisfies with the user specification (Program Verifier).
Whenever a program satisfies the examples, the synthesis process can terminate
and the program is translated into R. Finally, the translation layer produces
the SQL query from the R program. Next, each module is described in more detail.

\paragraph{Program Enumeration} \label{sec:squares-enum}
The purpose of the Program Enumerator is to continuously generate new candidate programs 
based on the specification. Programs are enumerated with the help of a \ac{SMT} solver, 
using a line-based representation~\cite{orvalhoCP19}. 
Programs are also enumerated in increasing number of DSL operations used.
Hence, if a limit on the number of DSL operations or a given time limit is reached
without being able to synthesize a verified program, then the tool ends with failure.

\paragraph{Program Verification and Translation}
In order to evaluate the candidate programs and check if they satisfy the user's 
specification, the synthesizer translates them into R. Then, the R program is executed 
using the input example that the user provided. The output is then compared with the 
expected output according to the example. This comparison treats tables as a multi-set 
of rows, meaning that row order is ignored. If the tables match, a solution to the 
problem has been found.
Finally, before presenting the program to the user, it must be translated into \ac{SQL}. 
To do this the \texttt{dbplyr} library is used. This library allows one to use regular 
databases as a back-end for \texttt{dplyr} operations and extract the corresponding 
\ac{SQL} queries.

\subsection{Sequential Synthesis in \cubes}
\label{sec:cubes_seq}

\cubes sequential engine improves on previous synthesizers~\cite{orvalhoVLDB20} by \emph{extending the domain specific language} to handle more expressive queries and by using \emph{pruning techniques} that prevent incorrect programs from being enumerated. From now on, we refer to our sequential query synthesizer as \cubesseq.

\paragraph{Extending the Domain Specific Language} 
\label{sec:cubes-dsl}
In order to support a wider range of programs, \cubes uses a more extensive and expressive \ac{DSL}. Observe that a larger \ac{DSL} allows more programs to be synthesized, but it also results in a larger search space.
Hence, some synthesizers focus on a smaller \ac{DSL} to target a narrower set of programs. However, \cubes tries to be more general and the growth of the \ac{DSL} is offset by the scalability gains of using new pruning techniques and parallel approaches.

The \ac{DSL} used in \cubes is detailed in \autoref{fig:cubes-dsl}.
\begin{figure}[t]
    \input{figures/cubes-dsl.tex}
    \caption{\ac{DSL} used by the \textsc{Cubes} synthesizer. 
    New components are highlighted in \texttt{\textbf{bold}}. 
    \label{fig:cubes-dsl}}
\end{figure}
Besides allowing more operations than \squares, \cubes also supports 
several new aggregation functions:
\texttt{n\_distinct}, \texttt{str\_count}, \texttt{cumsum}, \texttt{pmin}, \texttt{pmax}, \texttt{mode}, \texttt{lead}, \texttt{lag}, \texttt{median}, \texttt{rank} and \texttt{row\_number}.
Moreover, some aliases are also supported, in order to facilitate usage by users 
familiar with \ac{SQL}: \texttt{count} is an alias for both \texttt{n} and  
\texttt{n\_distinct} (activates both options), and \texttt{avg} is an alias for 
\texttt{mean}.
Finally, the type inference mechanism has been overhauled, resulting in dates and times now being supported. 

\paragraph{Pruning Invalid Programs} \label{sec:cubes-deducing}
One way to improve a program synthesizer is to reduce the number of incorrect programs that must be tested before finding a solution. In the case of \squares, a common example of such programs are those where at some point a column that does not exist in the current context is referenced. Consider the following program, which uses the tables from \autoref{fig:example} and the \ac{DSL} from \autoref{fig:cubes-dsl}:

\begin{lstlisting}{text}
df1 = filter(Courses, grade == 'A')
\end{lstlisting}

In this program, we are taking the \texttt{Courses} table and trying to filter its rows by selecting only the ones where column \texttt{grade} is equal to \texttt{`A'}. After a closer look at the \texttt{Courses} table it is clear that this program makes no sense, as there is no \texttt{grade} column in this table. \squares produces such candidate programs because it uses a \ac{DSL} that is defined at initialization time, containing all possible conditions, which is then given to the \ac{SMT} solver for it to generate candidate programs. Without some extra guidance, there is no way for the \ac{SMT} solver to only generate valid candidates.

We introduce a new form of pruning that eliminates these invalid programs. All component arguments are annotated with two sets of columns. These annotations are then used to constrain the set of programs that can be returned by the program enumerator, by implementing the column semantics of the corresponding \ac{DSL} operations. In Example~\ref{ex:filter-bitprune} we show how these annotations can be used to force all \texttt{filter} lines to always be valid. Note that the second annotation is only needed for some argument types, in order to record extra information such as in the case in Example~\ref{ex:mutate-bitprune}.

\begin{example} \label{ex:filter-bitprune}
  Consider again the previous program: \texttt{filter(Courses, grade == `A')}. The program contains just a \texttt{filter} operation that takes two arguments: \texttt{Courses} and \texttt{grade == `A'}.

  Arguments of type \(\mathit{table}\) are automatically annotated with the columns they contain, so in this case \texttt{Courses} would be annotated with \texttt{CourseId} and \texttt{CourseName}.
  In a similar fashion, we annotate filter condition arguments with the columns they require to be present to produce a valid program. In this case \texttt{grade == `A'} would be annotated with \(\texttt{grade}\). Finally, we encode that all filter operations must be such that all the columns in the annotation of the second argument appear in the annotation of the first argument to be valid.
  Since the presented program violates these rules it is deemed incorrect and is never generated by \cubes.
\end{example}

In order to propagate the column information along the several lines of the program, each line is also annotated with the set of columns available in the output table of that line. This information can then be used like that of any other argument of type \(\mathit{table}\). By implementing these kind of rules for all the components we can greatly reduce the number of enumerated programs that are invalid due to column names. As a result, the overall performance of \cubesseq is improved (as shown in \autoref{sec:evaluation}).

\autoref{fig:inference} shows the inference rules for all the components of our \ac{DSL}. These rules were obtained by carefully analyzing the documentation for the R functions used to implement the \ac{DSL} operations. Using these rules, we can infer from the arguments of a given operation what columns would be present in the output table if the line were executed. By extension, we can also determine invalid lines because no rule will be applicable to them. The contents of each annotation are described in \autoref{fig:inference-annot}.

\begin{figure}[t]
  \begin{gather*}
    \infer[\textsc{NaturalJoin}]{output' = table_1' \cup table_2'}{} \qquad
    \infer[\textsc{Filter}]{output' = table'}{filterCondition' \subseteq table'} \\
    \infer[\textsc{NaturalJoin3}]{output' = table_1' \cup table_2' \cup table_3'}{} \\
    \infer[\textsc{NaturalJoin4}]{output' = table_1' \cup table_2' \cup table_3' \cup table_4'}{} \\
    \infer[\textsc{InnerJoin}]{output' = table_1' \cup table_2'}{joinCondition' \subseteq table_1' & joinCondition'' \subseteq table_2'} \\
    \infer[\textsc{AntiJoin}]{output' = table_1'}{cols' \subseteq table_1' & cols' \subseteq table_2' & (cols' \neq \emptyset \lor table_1' \cap table_2' \neq \emptyset)} \\
    \infer[\textsc{LeftJoin}]{output' = table_1' \cup table_2'}{table_1' \cap table_2' \neq \emptyset} \qquad
    \infer[\textsc{Union}]{output' = table_1' \cup table_2'}{} \\
    \infer[\textsc{Intersect}]{output' = col'}{col' \subseteq table_1' & col' \subseteq table_2'} \qquad
    \infer[\textsc{SemiJoin}]{output' = table_1'}{table_1' \cap table_2' \neq \emptyset} \\
    \infer[\textsc{CrossJoin}]{output' = table_1' \cup table_2'}{crossJoinCondition' \subseteq table_1' & crossJoinCondition'' \subseteq (table_1' \cap table_2')} \\
    \infer[\textsc{Summarise}]{output' = summariseCondition'' \cup cols'}{
    \begin{gathered}
    summariseCondition' \subseteq table_1' \qquad cols' \subseteq table_1' \\[-1ex]
    (cols' \cap summariseCondition'') = \emptyset
    \end{gathered}
    } \\
    \infer[\textsc{Mutate}]{output' = table' \cup summariseCondition''}{summariseCondition' \subseteq table'}
  \end{gather*}
    \caption{Inference rules used to determine valid programs. \(A'\) denotes the first annotation of element \(A\), while \(A''\) denotes the second annotation. Where not mentioned, it is assumed that the second annotation is \(= \emptyset\). \label{fig:inference}}
\end{figure}

\begin{figure}[t]
  \begin{align*}
    table':&\; \text{columns present in the table} \\
    col':&\; \text{column required in the table} \\
    cols':&\; \text{columns required in the table} \\
    filterCondition':&\; \text{columns used in the filter condition} \\
    joinCondition':&\; \text{columns required in the first table} \\
    joinCondition'':&\; \text{columns required in the second table} \\
    crossJoinCondition':&\; \text{columns required in the first table} \\
    crossJoinCondition'':&\; \text{columns required in both tables} \\
    summariseCondition':&\; \text{columns used in the summarise condition} \\
    summariseCondition'':&\; \text{columns generated by the summarise condition}
  \end{align*}
  \caption{Semantics of each annotation.  \(A'\) denotes the first annotation of \(A\) 
  and \(A''\) denotes the second annotation. \label{fig:inference-annot}}
\end{figure}

\begin{example} \label{ex:mutate-bitprune}
  Consider the following \textit{summariseCondition}: \texttt{max\-}\texttt{StudentID = max(StudentID)}. The first annotation of a \textit{summariseCondition} corresponds to the columns that are ``used'', that is, the columns that must be present in order for the condition to be applicable. In this case the first annotation would be \(\texttt{StudentID}\). The second annotation corresponds to the columns that are generated by the \textit{summariseCondition}, in this case: \(\texttt{maxStudentID}\).
  
  When this condition is used in a \texttt{mutate}, rule \textsc{Mutate} from \autoref{fig:inference} states that if all of the required columns (first annotation) are present in the input, then we can conclude that the output table will be comprised of all columns that were already present in the input table, along with the generated columns (second annotation).
\end{example}

The two annotations of sets of columns are represented in \ac{SMT} as bit-vector variables, while the rules in \autoref{fig:inference} are implemented directly as constraints over those variables.
Hence, the corresponding invalid programs are never generated.

\subsection{Sequential Synthesis in other Synthesizers}

\subsubsection{\scythe}
\scythe \cite{scythe} is a sketch-based SQL synthesis tool. The search for a correct program occurs in two phases. In the first phase, \scythe enumerates all sketches (named \emph{abstract queries} in \scythe) of a given size. Next, \scythe evaluates these abstract queries. The execution of an abstract query is defined so that it contains all rows that could be present in the concrete instantiations of that abstract query. For example, for a \texttt{WHERE \(\square\)} operation, \scythe executes the operation as if the missing predicate were simply \texttt{TRUE}. \scythe discards all abstract queries that are not supersets of the output table (since no instantiation of those abstract queries will lead to a correct program). Finally, the synthesizer tries to fill all the missing predicates in the abstract query such that the output table is obtained, using techniques such as \emph{predicate equivalency} in order to speed up the process. If no correct, complete program is found, the depth is increased, and new abstract queries are enumerated.  By default, \scythe enumerates multiple solutions, ordered by the likelihood that they satisfy the user's intent.

\subsubsection{\patsql}
\patsql \cite{patsql} is a sketch-based synthesizer, like \scythe. The enumeration begins with a single \texttt{SELECT} statement or with \texttt{SELECT} and \texttt{ORDER BY} statements. This initial sketch is refined iteratively until a complete query is produced. This query is then executed, and its output compared with the one provided by the user. \patsql improves upon the previous state of the art by introducing new propagation algorithms for each of the supported operations, which allows it to propagate information from the output table throughout the query. Furthermore, the authors also modified the enumeration algorithm to generate only queries in normal form, effectively removing redundant queries from the program space.
Although \patsql supports enumerating more than one solution, it is designed to output a single solution, and the performance of the tool deteriorates if the user asks for multiple solutions~\cite{patsql}. 

\subsection{Synthesizer Comparison}

\begin{table}[t]
\caption{Input requirements for the different synthesizers.}
\label{tab:input-reqs}
\begin{minipage}{\columnwidth}
\center
\begin{tabular}{@{}lllll@{}}
\toprule
 & \scythe & \squares & \patsql & \cubes \\ \midrule
Constants & Yes & Yes & Yes & Yes \\
Comparison Columns\footnote{Columns which are used in comparisons with constants or other columns.} & No & Yes & No & Yes \\
Aggreg. Functions & Yes & Yes & No & Yes \\
Data Types & No & No & Yes & Optional\footnote{\cubes uses data types but it can also deduce them from the data, if not provided.} \\ \bottomrule
\end{tabular}
\end{minipage}
\end{table}

\begin{table}[t]
\caption{Comparison of features of the different synthesizers.}
\label{tab:synth-feats}
\begin{minipage}{\columnwidth}
\center
\begin{tabular}{@{}lllll@{}}
\toprule
 & \scythe & \squares & \patsql   & \cubes    \\ \midrule
\texttt{UNION} & Yes     & Yes     & No       & Yes      \\
\texttt{INTERSECT} & No & Yes & No & Yes      \\
Column Renaming & Yes & No & Yes & Yes      \\
\texttt{WINDOW} Functions\footnote {Operations such as cumulative sum, lag, lead, etc.} & No & No & Yes & Yes \\
Join w/ Condition & Yes & No & Yes & Yes \\   
Date/Time Arithmetic & No & No & Yes & No \\
\texttt{ORDER BY} & No & No & Yes\footnote{\label{foot:orderby}Will output an \texttt{ORDER BY} if it detects the output table is ordered.} & Yes\textsuperscript{\ref{foot:orderby}} \\ \bottomrule
\end{tabular}
\end{minipage}
\end{table}

The different SQL synthesizer tools presented have different capabilities and limitations, so a user will need to choose the one that best suits his needs. Next, we analyze several features of each tool, comparing them with \cubes.

\paragraph{Input requirements} Table~\ref{tab:input-reqs} shows the different input requirements for each tool. All tools require the user to provide any constants that the query might use. Both \squares and \cubes require the user to provide columns that are used in comparisons with constants or other columns.  \patsql is the only tool that does not require the user to provide aggregate functions (e.g., average) when the synthesized query must use aggregate functions. On the other hand, \patsql requires the user to provide the data types for each column.
These input requirements are used by the synthesis tools to prune the search space and make the synthesis process feasible. 
Even though all tools have some input requirements, they are usually easy for the user to specify and are related to his intention.

\paragraph{Query features} Table~\ref{tab:synth-feats} shows the different features of SQL supported by each SQL synthesis tool. Note that all of the SQL synthesizers presented can synthesize complex SQL queries that contain features such as natural or left joins, grouping and aggregation functions, filters, and compound filters, among others. They can also synthesize nested queries and are much more powerful than the first versions of SQL synthesis tools~\cite{DBLP:conf/sigmod/TranCP09,DBLP:conf/kbse/ZhangS13}.
We can observe that \cubes is one of the most full-featured synthesizers, supporting union and intersection operations, column renaming aggregation/window functions, complex joins and \texttt{ORDER BY} operations. Date and time arithmetic is one of the few features supported by other PBE SQL synthesizers (in this case \patsql) that is not supported by \cubes.

\paragraph{Multiple solutions} Examples are clearly an under-specification of the problem to be solved. It may be the case that the returned query satisfies the example provided by the user, but does not match his intention. Therefore, it may be of interest to the user to receive a ranked list of solutions instead of only a single query.

\squares does not support returning multiple solutions. On the other hand, \scythe returns by default up to 5 queries that match the example provided by the user. This limit can be changed by modifying the source code, but requires increasing the beam size used during exploration, which has a significant impact on the performance of the synthesizer. In practice, asking for a large number of solutions is not feasible with \scythe and the user should use the default options. 
\patsql can also be used with the option to request multiple solutions. When this option is used, \patsql attempts to find alternative queries that satisfy the same example. 
However, this procedure degrades the performance of the tool and the authors do not suggest that this option be used~\cite{patsql}.
\cubes can find multiple solutions by continuing the search until a given time limit is reached or a maximum number of solutions is found. 
All solutions found within the resource limit are provided in the output.
The overhead of this option is that, instead of terminating as soon as a query that satisfies the example is found, it runs until the resource limit is reached.

\section{Parallel Synthesis}
\label{sec:parallel}

In this section, we propose \cubesdcB, a divide-and-conquer approach for \cubes where the 
synthesis problem is split into several disjoint sub-problems to be solved in parallel.
Our strategy is to split the problem into many smaller sub-problems that are 
solved independently by each of the processes. 
Hence, each process focuses solely on a particular area of the search space.

\begin{figure}[t]
\begin{center}
    \resizebox{.9\linewidth}{!}{
\begin{tikzpicture}

  \node (sp) at (0.5, 2) {Spec.};

  \node[box1] (cg) at (2.5, 2) [right = .6cm of sp, text width=1.9cm] {Cube Generator};

  \node[box2] (c3) [right = .6cm of cg] {Cube 3};
  \node[box2] (c2) [above = .5cm of c3] {Cube 2};
  \node[box2] (c1) [above = .5cm of c2] {Cube 1};
  \node[tef1] (dots) [below = .1cm of c3] {\vdots};
  \node[box2] (cn) [below = .25cm of dots] {Cube n};

  \draw[->]   (cg.east)  -- (c1.west);
  \draw[->]   (cg.east)  -- (c2.west);
  \draw[->]   (cg.east)  -- (c3.west);
  \draw[->]   (cg.east)  -- (cn.west);

  \node[box1d] (synt1) [right = .6cm of c1] {Synthesis};
  \node[box1d] (synt2) [right = .6cm of c2] {Synthesis};
  \node[box1d] (synt3) [right = .6cm of c3] {Synthesis};
  \node[tef1]  (sdots) [below = .1cm of synt3] {\vdots};
  \node[box1d] (syntn) [right = .6cm of cn] {Synthesis};

  \draw[->]   (c1.east)  -- (synt1.west);
  \draw[->]   (c2.east)  -- (synt2.west);
  \draw[->]   (c3.east)  -- (synt3.west);
  \draw[->]   (cn.east)  -- (syntn.west);

  \node[box1] (tl) [right = .6cm of synt3, text width=1.65cm] {Translation Layer};

  \draw[->]   (synt1.east)  -- (tl.north west)
  node[midway, right, xshift=.5em, align=left, text width=5em] (l2)
  {R Program\\or Fail};
  \draw[->]   (synt2.east)  -- ([yshift=-1em]tl.north west);
  \draw[->]   (synt3.east)  -- (tl.west);
  \draw[->]   (syntn.east)  -- (tl.south west);

  \node[draw,dashed,fit=(cg) (dots) (synt1) (syntn) (tl)] (cubes) {};

  \node[tef1] (prog) [right = .5cm of cubes, yshift=.5cm, anchor=south west, text width=1.4cm, align=left] {SQL};
  \node[tef1] (fail) [right = .5cm of cubes, yshift=-.5cm, anchor=north west, text width=1.4cm, align=left] {Fail};

  \node[tef1] (cubes-label) [above = 0cm of cubes.north east, anchor=south east, text width=1.9cm, align=right] {\cubesdcB};

  \draw[->]   (sp)  -- (cubes);
  \draw[->]   (cubes.east) -- (prog.south west);
  \draw[->]   (cubes.east) -- (fail.north west);

\end{tikzpicture}}
    \end{center}
    \caption{\cubes' architecture for divide-and-conquer.} \label{fig:cubes-dc}
\end{figure}
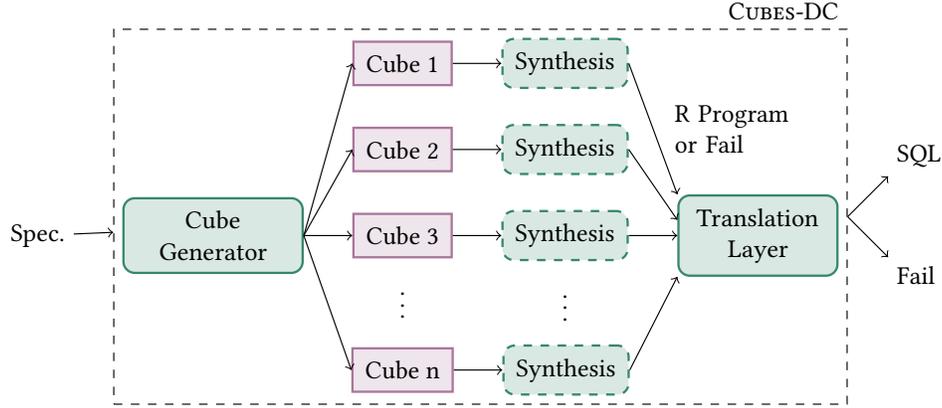

In our context, each sub-problem is represented by a \emph{cube}: a sequence of operations from the \ac{DSL} such that the arguments for the operations are still to be determined. A \emph{cube} can also been seen as a special type of sketch where all the operations have been filled and only terminal symbols are missing.
Consider the following cube as an example: \texttt{[filter, natural\_join]}, which represents the section of the search space composed by programs with two operations, where the first is a \texttt{filter} and the second is a \texttt{natural\_join}.

Each process receives a specific cube and checks if it is possible to extend it to a complete program that satisfies the input-output example. If this is not possible, then the cube is deemed unsatisfiable and the process requests a new cube to explore.
Observe that each cube corresponds to a distinct sequence of operations, and as such, 
there is no intersection in the search space of each process.
The overall architecture is illustrated in \autoref{fig:cubes-dc}. The cube
generator component is responsible to generate cubes in increasing size (e.g.
first the cubes with one operation, then with two operations, and so forth).
Whenever a process finds a solution, then the translation layer transforms 
the DSL program into SQL.
Otherwise, if a cube cannot be extended into a complete program that satisfies 
the user specification, then the process gets a new cube from the queue.

\paragraph{Dynamic Cube Generation.}
One approach for a cube generation heuristic is to define a static order
of operations to be explored. Although a static heuristic can be effective on
some specific domains, it is very unlikely that it generalizes to new
instances. 
Therefore, \cubesdcB uses a dynamic cube generator inspired on \ac{NLP} techniques.
Since candidate programs are constructed as a sequence of operations, a bigram
prediction model can be used to decide the operation to be chosen next in a 
given sequence.
Therefore, when choosing the operation in a given sequence, the operation immediately 
preceding it is used to compute an expectation of which of the possible choices 
will lead to the desired program.
Hence, for each pair of operations (\texttt{a}, \texttt{b}) a score 
\(S_{\texttt{a},\texttt{b}}\) is maintained. The value of \(S_{\texttt{a},\texttt{b}}\)  
represents the expectation that using a \texttt{b} operation after an 
\texttt{a} operation will lead to the desired program. Scores are 
updated as programs are evaluated, as explained next.

\paragraph{Program scoring.} For a given program \(p\), let \texttt{output}
denote the result of executing program \(p\) on a given example specified by
the user. Moreover, let \texttt{expected} denote the desired result provided in
the input-output example.
First, we compute the set of all values in the \texttt{output} and
\texttt{expected} tables: \(\mathbf{unique}(\texttt{output})\) and
\(\mathbf{unique}(\texttt{expected})\).
Next, we compute the score of program \(p\) as the percentage of elements of the
expected output that appear in the output of \(p\) as follows:
\begin{align}
score(p) =& \frac{|\mathbf{unique}(\texttt{output})\,\cap\,\mathbf{unique}(\texttt{expected})|}{|\mathbf{unique}(\texttt{expected})|}
\end{align}

A score of 1 indicates that all the expected values occur in the output, and as
such, a filtering or restructuring might lead to a correct program.
On the other hand, a value of 0 means that the candidate program is likely very
far from a correct solution. Note that any program \(p\) where \(score(p) \neq 1\) is 
certainly incorrect. This can be used as an optimization in order to avoid expensive 
table comparisons.

\paragraph{Score updates} For each evaluated program \(p\), the score, 
\(\mathit{score}(p)\), is used to update the bigram scores. For example, let
\(p\) denote a program that uses the following operations: 
\texttt{filter}, \texttt{natural\_join}, \texttt{summarise}  (in that order). 
Then, the scores for the bigrams that appear in the program are updated as follows:
    \begin{align}
        S_{\emptyset, \texttt{filter}} & \pluseq \mathit{score}(p) \\
        S_{\texttt{filter}, \texttt{natural\_join}} & \pluseq \mathit{score}(p) \\
        S_{\texttt{natural\_join}, \texttt{summarise}} & \pluseq \mathit{score}(p)
    \end{align}
Furthermore, the scores of the operations occurring in the beginning of the
sequence (including the first operation) are also updated, although with 
decreasing weights.
In particular, the operation selected for position \(i\) (zero-based) of
the sequence contributes with \(\frac{1}{(i+1)^2} \cdot \mathit{score}(p)\).
Hence, considering again the program \(p\) with components \texttt{filter},
\texttt{natural\_join}, and \texttt{summarise}, the updates are as follows:
\begin{align}
    S_{\emptyset, \texttt{filter}} & \pluseq \sfrac{1}{1} \cdot \mathit{score}(p) \\
    S_{\emptyset, \texttt{natural\_join}} & \pluseq \sfrac{1}{4} \cdot \mathit{score}(p) \\
    S_{\emptyset, \texttt{summarise}} & \pluseq \sfrac{1}{9} \cdot \mathit{score}(p)
\end{align}
These extra score updates are done in order to introduce a small chance of 
reordering the operations.

\paragraph{Cube construction.} 
Cubes are constructed by iteratively adding operations to a sequence up to a given
limit on the number of operations.
Suppose that the last selected operation is \texttt{op} (if no operation has been 
selected, i.e. the cube is empty, \texttt{op} is the empty symbol \(\emptyset\)).
In order to decide which operation should follow \texttt{op}, all the scores for that 
prefix, \(S_{\texttt{op},\textit{op}_1}, ..., S_{\texttt{op},\textit{op}_n}\), are 
retrieved, normalized and smoothed, using Laplace
smoothing~\cite{DBLP:books/lib/JurafskyM09}. These steps result in a list of probabilities
that corresponds to the expectation values associated with each operation.
The operation for the current line is then chosen from a distribution using those
expectation values. This is done until we have a cube of the desired size.
A compact tree structure is used to keep track of already generated cubes, as to avoid
repetition.

\paragraph{Avoiding biases.} The usage of the dynamic cube generation technique may
introduce biases since the bigram scores are continuously increasing. In particular, operations
that are selected first become more likely to be selected again when generating
new cubes.
Two methods are used to handle this issue:
\begin{itemize}
    \item Each time a new program is generated, all scores are multiplied by a
    $\delta$ such that $0 < \delta < 1$. By default, $\delta = 0.99999$ is used.
    This is done so that past information can be gradually forgotten, in order 
    to increase diversity in the exploration of the search space. These updates 
    are done in batches, in order to not overwhelm inter-process communication.
    \item A fixed number of processes, by default 2, always solve randomly
    generated cubes (as long as not previously generated), in order to diversify the
    search process.
\end{itemize}

\paragraph{DSL Splitting.} 
\label{sec:split}
Besides the splitting of the search space using cubes, \cubesdcB also includes
the splitting of the DSL operations among the processes. The motivation for
this additional split is that two of the \ac{DSL} components, \texttt{inner\_join} 
and \texttt{cross\_join},
are much more complex than any of the other operations. That is, there are many more 
ways to complete a \texttt{cross\_join} operation than, for example, a 
\texttt{summarise} operation. 
In fact, the difference in complexity is so large that building the SMT encoding 
becomes much more computationally expensive when those operations are 
included.
Therefore, the available processes in \cubesdcB are split into two sets: 
set $F$ only considers programs that contain at least one of these two operations, 
while set $B$ considers only the sub-space of programs that do not contain any
of these operations.

If the program to be synthesized does require one of the two complex joins, then 
the encoding overhead is unavoidable and one of the processes in set $F$ will 
lead to a solution.
On the other hand, if the desired program does not require a complex join, then the 
overhead is restricted to the processes in set $F$, while the processes in set $B$
avoid the overhead. 
The goal is then to balance the number of processes allocated to each set in 
order to maximize the programs that can be synthesized within a given time limit.
The ratio between the number of processes in sets $F$ and $B$ is configurable and 
defaults to 1:2.

\section{Accuracy and Disambiguation}
\label{sec:disambiguation}
An essential issue in program synthesis is knowing if the returned program corresponds to the user intent.
To determine the accuracy of the synthesis tools, we call the query that the user wishes to obtain the \emph{ground truth} query.
Observe that SQL synthesis tools that use input-output examples return a query that satisfies the user's examples.
However, these examples are an under-specification, and as such, the returned query might not satisfy the true user intent.  

\cubes may find multiple queries that satisfy the examples. However, unless these queries are equivalent, only one of them matches the user's intent. To address this challenge, we propose creating new examples with different input-output pairs for the possible synthesized queries and interacting with the user to disambiguate the correct query. This section discusses how we use fuzzing to create new examples and our disambiguation procedure to improve the accuracy of \cubes and meet the user intent.

\subsection{Fuzzing} \label{sec:fuzzing}
Given a set of queries returned by \cubes, we want to determine which one matches the user intent. Since some of them may be equivalent, multiple queries may be correct. One approach could be to use query equivalence tools to check the equivalence of these queries and only consider a representative query of each equivalence class.
Although recent work in query equivalence tools \cite{DBLP:conf/cidr/ChuWWC17, DBLP:journals/pvldb/ZhouANHX19, DBLP:journals/pvldb/ChuMRCS18} has advanced the state-of-the-art, these tools remain incomplete, not supporting many complex queries present in our datasets. To overcome this limitation, we use a fuzzing-based approach to determine the approximate equivalency of different queries.

Consider a synthesis problem with an input-output example \((I, O)\). Fuzzing consists in taking the input \(I\), slightly modifying it, producing \(I'\) and then running the ground truth with \(I'\) as an input. This produces a new input-output pair \((I', O')\) which can be used to verify if a given query is consistent with the ground truth. If the outputs of the two queries being checked for equivalency differ then the queries are surely distinct.
However, if the output is the same, we still cannot conclude that the queries are equivalent.
Hence, we perform several rounds of fuzzing, generating different input-output pairs, with each round increasing the confidence in our answer.

In order to produce fuzzed input-output examples, we use the Semantic Evaluation suite \cite{DBLP:conf/emnlp/ZhongYK20}. Consider a table, \(T \in I\). In order to generate a fuzzed version of this table, \(T' \in I'\), the suite starts by randomly selecting the number of rows that the new table will have. Then, in order to fill the cells of \(T'\), three sources are used: (1) values sampled from a uniform distribution for the given type (i.e., for integers a uniform distribution on \([-2^{63},\,2^{63}-1]\)), (2) values taken from the corresponding columns on the original table, \(T\), and closely related values (i.e., if ``Alice'' is in \(T\) then both ``Alice'' and ``Alicegg'' might be considered for \(T'\)), and (3) values taken from the queries we are comparing, and closely related values. The reason why the suite also takes into account values from the queries themselves is to increase code coverage (for example, making it more likely to find off-by-one errors). Furthermore, all foreign keys are respected so that the semantics of the database are preserved.

\subsection{Disambiguation}
\cubes is able to return multiple queries that satisfy the user specification. However, if the example provided is an under-specification of the true user intent, those queries will most likely have slightly different semantics. In order to ease the burden on the user of selecting a correct query, we propose a disambiguation algorithm, shown in Algorithm~\ref{alg:dis}.

\begin{algorithm}[t]
\caption{Disambiguation method}\label{alg:dis}
\SetKwFunction{Execute}{Execute}
\SetKwFunction{Fuzz}{Fuzz}
\SetKwFunction{Remove}{Remove}
\SetKwFunction{GroupByOutput}{GroupByOutput}
\SetKwFunction{BetterSplit}{BetterSplit}
\SetKwFunction{AskUserIfExampleIsCorrect}{AskUserIfExampleIsCorrect}
\SetKwFunction{First}{First}
\SetKwData{BestSplit}{bestSplit}
\SetKwData{Split}{split}
\KwIn{$\mathcal{S}$, the set of synthesized queries}
\myinput{\(I\), input database}
\myinput{\(O\), output table}
\myinput{\(R\), number of fuzzing rounds}
\KwResult{a query considered to be the most likely solution}

\SetKwFunction{Disambiguate}{Disambiguate}
\Indm\Disambiguate{\(\mathcal{S}, I, O, R\)}\\
\Indp

\nl\BestSplit \(\gets \emptyset\)\;

\nl\For{\(i \leftarrow 1\) \KwTo \(R\)}{
    \nl\(I' \gets \) \Fuzz{\(I\), \(\mathcal{S}\)}\; 
    
    \nl\Split \(\gets\) \GroupByOutput{\(\mathcal{S}\), \(I'\)}\;
    
    \nl\uIf{\BetterSplit{\BestSplit, \Split}}{
        \nl\BestSplit \(\gets\) \Split\;
    }
}

\nl\uIf{\BestSplit \(= \emptyset\)}{
    \nl\Return{\First{\(\mathcal{S}\)}}\label{algl:return_first}\;
}

\nl\((I', \mathcal{S}_A, O'_A, \mathcal{S}_B) \gets\) \BestSplit\;

\nl\uIf{\AskUserIfExampleIsCorrect{\(I'\), \(O'_A\)}}{
    \nl\Return{\Disambiguate{\(\mathcal{S}_A\), \(I\), \(O\), \(R\)}}\;
}
\nl\uElse{
    \nl\Return{\Disambiguate{\(\mathcal{S}_B\), \(I\), \(O\), \(R\)}}\;
}
\end{algorithm}

The disambiguation method used in \cubes starts by synthesizing all possible solutions under a given time limit. The goal is then to ask the user questions in order to iteratively discard queries until we are confident we found one that satisfies the user intent. In order to reduce the burden on the user we try to minimize the number of questions asked as much as possible, by trying to discard approximately half of the queries each time we ask a question. In Section~\ref{sec:evaluation} we show that this goal is verified in practice.

To do this, we start by generating a new input database \(I'\) through fuzzing. 
Then, 
we execute each of the synthesized queries on this new input \(I'\) and group them according 
to the output they produce. In each disambiguation step, we generate 16 new input databases, 
by performing fuzzing 16 times, and select the input-output example that is closest to 
splitting the set of queries in half. 

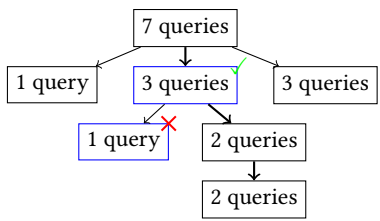
\begin{figure}
    \centering
    \newcommand{\Cross}{$\mathbin{\tikz [x=1.4ex,y=1.4ex,line width=.2ex, red] \draw (0,0) -- (1,1) (0,1) -- (1,0);}$}%

\newcommand{\Checkmark}{$\color{green}\checkmark$}

\begin{tikzpicture}[node distance=2em]

\node[draw] (P1) {7 queries};

\node[draw] (O11) [below left = 0.8em and 1.5em of P1] {1 query};
\node[draw=blue] (O12) [below = 0.8em of P1] {3 queries};
\node[draw] (O13) [below right = 0.8em and 1.5em of P1] {3 queries};
\node       (O1R) [above right = -1em of O12] {\Checkmark};

\node[draw=blue] (O22) [below left = 0.8em and -1.5em of O12] {1 query};
\node[draw] (O23) [below right = 0.8em and -1.5em of O12] {2 queries};
\node       (O2R) [above right = -1em of O22] {\Cross};

\node[draw] (O31) [below = 0.8em of O23] {2 queries};

\draw[->] (P1) -- (O11);
\draw[->, thick] (P1) -- (O12);
\draw[->] (P1) -- (O13);

\draw[->] (O12) -- (O22);
\draw[->, thick] (O12) -- (O23);

\draw[->, thick] (O23) -- (O31);

\end{tikzpicture}













    \caption{Example disambiguation process from a problem that generated 7 possible queries. Blue boxes represent the input-output example given to the user.}
    \label{fig:disambiguate}
\end{figure}

\autoref{fig:disambiguate} shows a real-world disambiguation interaction. Initially
we have 7 queries found by \cubes that satisfy the original input-output example.
In this case, we are able to generate a new input \(I'\) such that 1 of
the 7 queries provides the output table $A'$, 3 queries provide another output
table $B'$, and 3 other provide an output $C'$. Then, we ask the user if the new input-output example \((I', B')\)
is correct. If the user answers yes, then the solution is one of the 3 queries.
Otherwise, the solution should be one of the 4 remaining queries. Since the user answered yes,
then 3 queries remain to disambiguate.
The disambiguation procedure terminates when either 
there is only one query remaining or the fuzzing procedure is unable to find a new
example to distinguish the remaining queries. In the latter case, the remaining queries
are deemed equivalent and the first one found by \cubes during the search is returned 
to the user. Notice that \cubes enumerates queries in increasing order of the number 
of operators. 
Hence, the first queries to be found by \cubes have the fewest operations and
should be more general.

\section{Methods and Data} \label{sec:method}
In this section we describe the different benchmarks used to evaluate \cubes and compare it to other synthesizers, as well as the methods to perform that comparison.

\subsection{Data} \label{sec:data}

\begin{table}[t]
\caption{Overview of the benchmark sets used in the analysis.}
\label{tab:benchmarks}
\begin{tabular}{@{}lr@{}}
\toprule
 & Number of Instances \\ \midrule
\texttt{recent-posts} & 51 \\
\texttt{top-rated-posts} & 57 \\
\texttt{textbook} & 35 \\
\texttt{spider} & 3690 \\   
\texttt{kaggle} & 33 \\ \midrule
Total & 3866 \\ \bottomrule
\end{tabular}
\end{table}

We use five different benchmark sets, divided into two groups. The first group, consisting of the benchmark sets \rposts, \trposts, \textbook and \kaggle refers to benchmarks that were previously used in other example-based SQL synthesis papers. The second group consists of a single benchmark set: \spider. We adapted the instances in \spider from a very large and diverse dataset of queries used for SQL synthesis from \ac{NL} descriptions (also known as text-to-SQL) \cite{DBLP:conf/emnlp/YuZYYWLMLYRZR18}. The number of instances in each benchmark set is shown in \autoref{tab:benchmarks}.

The \(51+57\) instances from \rposts and \trposts were collected from questions asked in the StackOverflow\footnote{\url{https://stackoverflow.com/}} website. These instances were first used to evaluate the synthesizer \scythe \cite{scythe}. The 35 instances in the \textbook benchmark were collected from exercises in Sections 5.1 to 5.3 of the Database Management Systems book \cite{textbook}. This set has been used to evaluate most previous example-based SQL synthesizers \cite{scythe,DBLP:conf/kbse/ZhangS13, squares,patsql}. The 33 instances in the \kaggle benchmark set were extracted from Kaggle\footnote{\url{https://www.kaggle.com/}} SQL tutorials and were first used to evaluate \patsql \cite{patsql}. Finally, the 3690 instances in the \spider set were obtained by transforming the original Spider dataset \cite{DBLP:conf/emnlp/YuZYYWLMLYRZR18}. Spider was originally a data set to evaluate SQL synthesizers based on \ac{NL} descriptions, instead of input-output examples. For our purpose, we discarded the \ac{NL} descriptions and executed the ground truth queries over the sample databases, resulting in the output table for the I/O example. Furthermore, we used an extractor to obtain the rest of the information required by the different synthesizers (see \autoref{tab:input-reqs}) directly from the ground truth query. We discarded all duplicated queries, as well as all queries that were not executable using SQLite or that produced an output table with no rows.

\begin{figure}[t]
    \centering
    \includegraphics{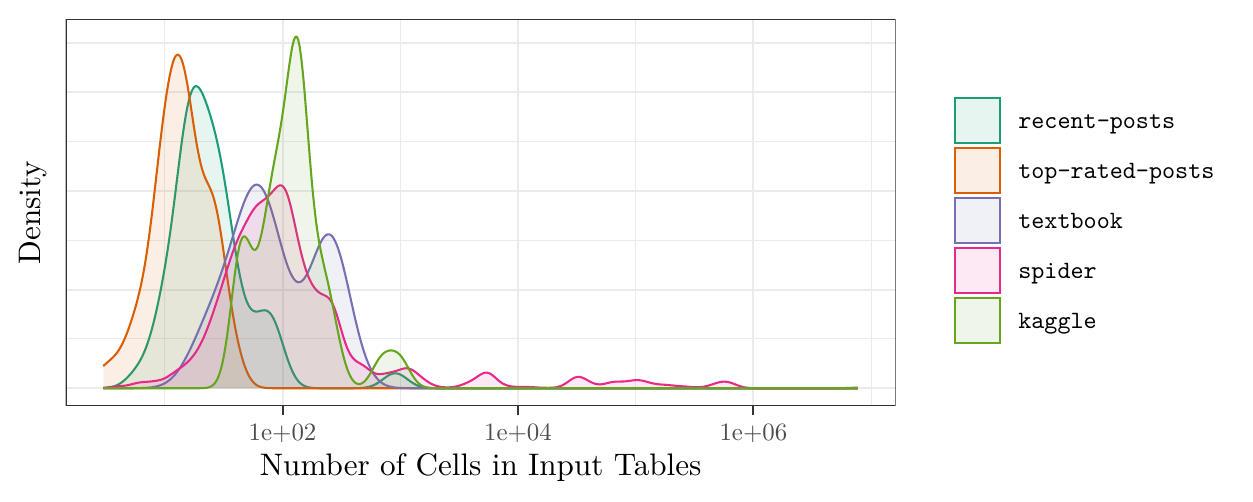}
    \caption{Total number of cells in input tables.}
    \label{fig:cells}
\end{figure}

\autoref{fig:cells} shows the distribution of the number of total cells in the input tables of the instances, grouped by the benchmark to which the instances belong. The density functions in these plots can be seen as an alternative to histograms that make it easier to read and compare the different benchmark sets\footnote{The density functions were obtained through \ac{KDE}.}. The plot shows that \rposts and \trposts have the lowest number of cells, on average, followed by \textbook and \spider and finally \kaggle with the largest average input database. However, we can also see that \spider has some very large instances with \(1000\) to more than \(10^6\) cells. It is expected that most synthesizers will have a harder time solving these instances, since at the very least they need to read the tables and execute the final query over those tables.

\begin{figure}[t]
    \centering
    \includegraphics{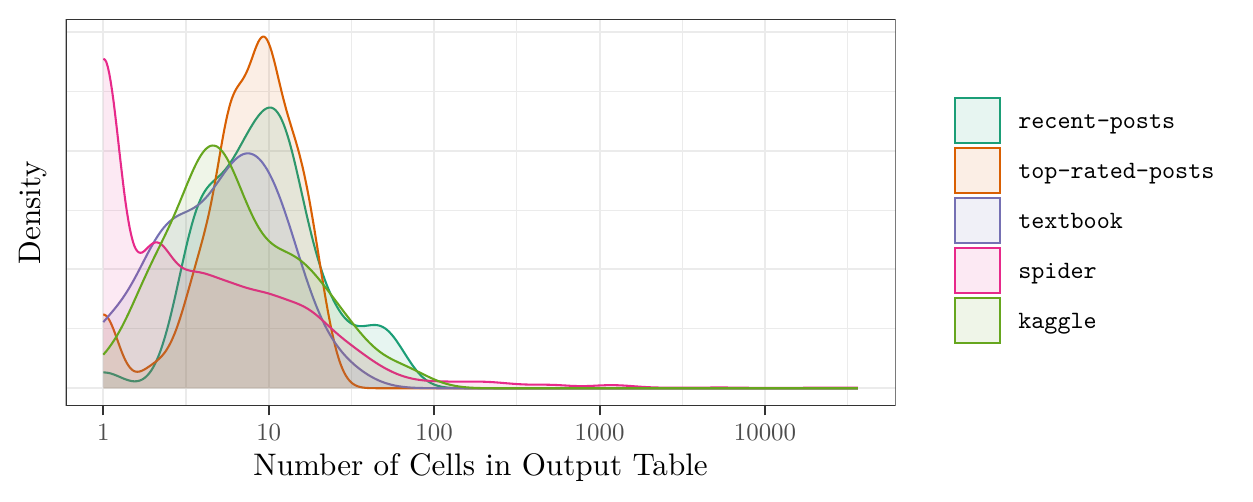}
    \caption{Total number of cells in the output table.}
    \label{fig:output_cells}
\end{figure}

\autoref{fig:output_cells} presents a similar plot, with the distribution of the number of cells in the output table for each instance, grouped by benchmark set. Here we can see that all sets except \spider have similar density functions with around 10 cells on average. Meanwhile, a very large number of instances in \spider have just one cell in the output table. There are two possible explanations for this: (1) the queries in question are very simple, for example, just summing all rows in a given column, or (2) these examples are under-specified; for example, the queries correspond to complex filter operations but there is not enough information in the input tables to produce more than one row in the output.

\begin{figure}[t]
    \centering
    \includegraphics{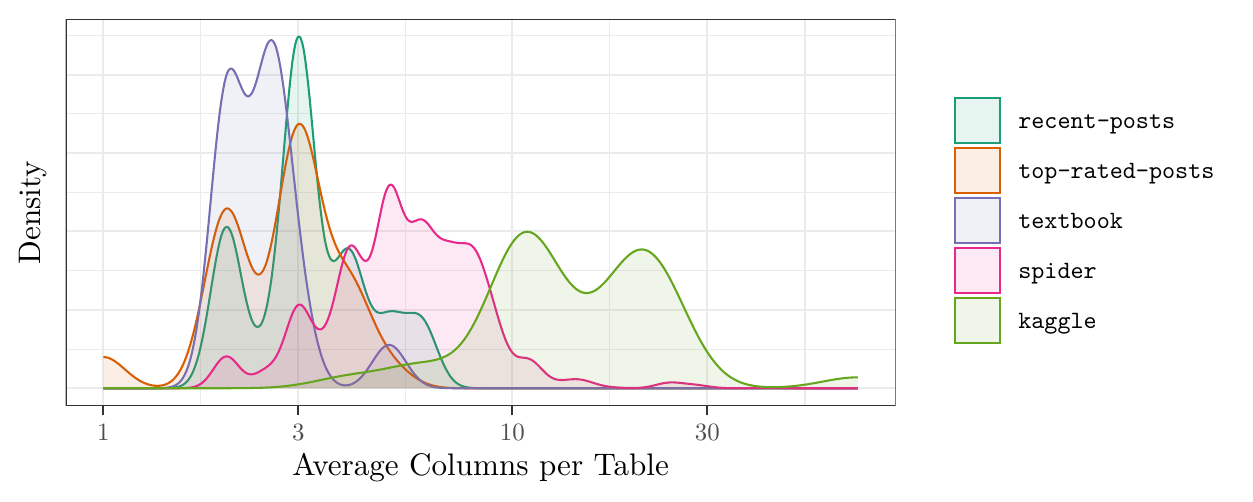}
    \caption{Average number of columns per input table}
    \label{fig:cols_per_table}
\end{figure}

\autoref{fig:cols_per_table} shows the distribution of the average number of columns per table, that is, the total number of columns in the input tables divided by the number of input tables. This is an interesting metric because it is directly correlated with the size of the program space that needs to be explored for each instance --- more columns means more possible combinations for each SQL operation. We can see that \rposts, \trposts and \textbook have similar distributions with around 4.2 columns per table on average. Meanwhile, the \spider benchmark set appears to be slightly more complex, with around 8.7 columns per table on average. By contrast, \kaggle has a much higher average number of columns per input table -- 21.3. This might mean that synthesizers will struggle with solving instances from this set.

\subsection{Methods}
In order to evaluate \cubes and compare it to other SQL synthesizers we use two distinct evaluation methods: simple evaluation and fuzzy-based evaluation.

\subsubsection{Simple Evaluation}
In this setting, we are simply interested in checking if a synthesizer can produce a query that satisfies the specification given by the user. That is, when executed, the query should produce an output table that is equal to the one specified by the user. Furthermore, we do not take into account the row order of the output table. This method has been extensively used in the past to measure the performance of SQL synthesizers \cite{scythe,DBLP:conf/kbse/ZhangS13,squares,patsql}. The problem with simple evaluation is that, in the case of an ambiguous example, it does not address whether the synthesized query actually satisfies the user intent or not.

\subsubsection{Fuzzy-based Evaluation}
In this setting, we are instead checking if the synthesized queries satisfy the true intent of the user, and not just the specification (input-output example). The motive for this distinction is that the input-output example might be an under-specification of the query the user wishes to obtain. That is, there are many queries that satisfy the example but not all of them have the same semantics.

\begin{algorithm}[t]
\caption{Query checker using fuzzing}\label{alg:fuzzy-eq}
\SetKwFunction{Execute}{Execute}
\SetKwFunction{Fuzz}{Fuzz}
\KwIn{$q$, the synthesized query}
\myinput{\(Q\), the ground truth query}
\myinput{\(I\), input database}
\myinput{\(R\), number of fuzzing rounds}
\KwResult{a Boolean representing if the algorithm was unable to find a distinguishing input}

\SetKwFunction{FuzzyCheck}{FuzzyCheck}
\Indm\FuzzyCheck{\(q, Q, I, R\)}\\
\Indp
\nl\uIf{\Execute{\(Q\), \(I\)} \(\neq\) \Execute{\(q\), \(I\)}}{
    \nl\Return{False}\; 
}

\nl\For{\(i \leftarrow 1\) \KwTo \(R\)}{
    \nl\(I' \gets \) \Fuzz{\(I\), \(Q\)}\; 
    \nl\uIf{\Execute{\(Q\), \(I'\)} \(\neq\) \Execute{\(q\), \(I'\)}}{
        \nl\Return{False}\;
    }
}

\nl\Return{True}\;
\end{algorithm}

Algorithm~\ref{alg:fuzzy-eq} shows how we use fuzzing, as introduced in \autoref{sec:fuzzing}, to determine if two queries are likely to have the same semantics. We start by sanity checking if the synthesized query, \(q\), and the ground truth query, \(Q\), produce the same output for the provided input database, \(I\) (lines 1 and 2). Then, we perform \(R\) rounds of fuzzing (line 3), where for each round, we generate a new input database, \(I'\), and check if the two queries still produce the same output table (lines 5 and 6). If all rounds pass successfully, we consider the queries equivalent (line 7). When comparing two tables, we perform a very lax comparison that: (1) ignores row order -- tables are seen as a multiset of rows, (2) ignores column names, and (3) tries to convert the datatypes of columns -- if two columns contain the same data but one as a number and the other as a string, they are considered equivalent. Finally, as with any fuzzing-based approach, several rounds might be needed to find an input that distinguishes the queries. Parameter \(R\) controls the maximum number of fuzzing rounds until the algorithm deems the queries equivalent.

\section{Evaluation}
\label{sec:evaluation}

The experimental results presented in this section aim to answer the following 
research questions:
\begin{enumerate}[label=\textbf{{Q}\arabic{*}.}]
\item How does \cubesseq compare with
other state-of-the-art SQL synthesizers when using the simple evaluation metric? (\autoref{subsec:sequential})
\item What are the speedups obtained by \cubesdc, when using the simple evaluation metric? (\autoref{subsec:parallel})
\item How does the structure of the input-output examples impact the performance of \cubes? (\autoref{subsec:io-structure-results})
\item How do \cubes and the other SQL synthesizers perform when using the fuzzy-based evaluation metric? (\autoref{subsec:accuracy})
\item What is the impact of program disambiguation in \cubes' fuzzy-based evaluation metric? (\autoref{subsec:accuracy})
\item What is the impact of non-determinism in \cubesdc? (\autoref{sec:non-determinism})
\end{enumerate}

All results were obtained on a dual socket Intel\textsuperscript{\textregistered} 
Xeon\textsuperscript{\textregistered} Silver 4210R @ 2.40GHz, with a total of 20 cores
and 64GB of RAM. Furthermore, a limit of 10 minutes (wall-clock time) and 56GB 
of RAM was imposed on all synthesizers (sequential or parallel). 
All limits were strictly imposed using \texttt{runsolver}~\cite{runsolver}.

\subsection{Implementation}
\cubes is implemented on top of the Trinity \cite{trinity} framework, using Python 3.8.3.
Candidate programs are enumerated using the Z3 \cite{z3} \ac{SMT} solver (version 4.8.8) with the Quantifier Free Finite Domain theory enabled. This theory transforms the generated \ac{SMT} formula into an equivalent \ac{SAT} formula which is then solved using an integrated \ac{SAT} solver.

Candidate programs are evaluated by translating the DSL operations into equivalent R instructions. In particular, the \texttt{tidyverse}\footnote{\url{https://www.tidyverse.org/}} family of packages is used to implement table manipulation instructions. Once a correct R program is found, the \texttt{dbplyr}\footnote{\url{https://dbplyr.tidyverse.org/}} package (version 1.4.4) is used to translate that program to an equivalent \ac{SQL} query.
In the parallel synthesizer inter-process 
communication is achieved using a message passing approach through Python's 
\texttt{multiprocessing} pipes. 

We use the fuzzing framework developed by \citet{DBLP:conf/emnlp/ZhongYK20} in our disambiguation module and to perform accuracy analysis. Furthermore, queries are executed using the SQLAlchemy\footnote{\url{https://www.sqlalchemy.org/}} library (version 1.3.20) and row order is ignored when comparing tables. In the original implementation, the fuzzing framework was non-deterministic so we modified it in two important ways: (1) we added proper seeding for Python's pseudo-random number generator, which is used in the tool, and (2) we replaced all usages of the \texttt{set} data structure with \texttt{OrderedSet} (sets backed with a list so that the iteration order is deterministic). This change was needed so that both the accuracy results presented in this paper and \cubes' disambiguation process were reproducible. The modified framework is also included in \cubes' source files\footnote{\url{https://github.com/OutSystems/CUBES}}.

\subsection{Sequential Performance using Simple Evaluation}
\label{subsec:sequential}

\begin{figure}[t]
    \centering
    \includegraphics{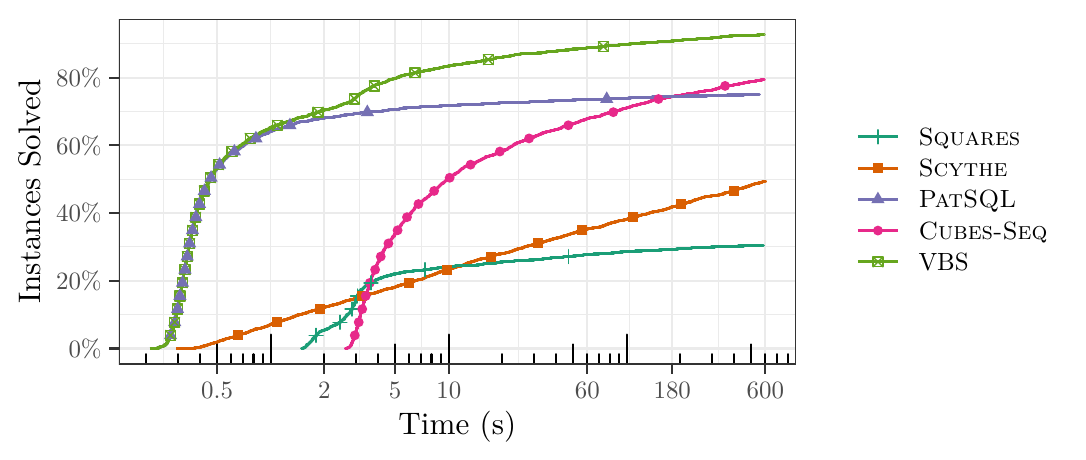}
    \vspace{-2mm}
    \caption{Percentage of instances solved by each tool at each point in time. A mark is placed every 150 solved instances.} \label{fig:seq-solved}
\end{figure}

In this section we evaluate the performance of \cubesseq, the sequential version of 
\cubes, and perform a comparison with other state-of-the-art  \ac{SQL} \ac{PBE} tools:
\squares~\cite{orvalhoVLDB20}, \scythe~\cite{scythe} and \patsql~\cite{patsql}.
In \autoref{fig:seq-solved}, we show the percentage of instances solved by each synthesizer as a function of time when using the simple evaluation method.
Overall, \squares was able to solve 30.6\% of the instances within the time limit
of 10 minutes, while \textsc{Scythe} solved 49.5\% and \patsql solved 75.1\%. 
\cubesseq was able to solve 79.4\%.

\autoref{fig:seq-solved} also shows the \ac{VBS} for these four synthesizers. The \ac{VBS} can be seen as the result of running the four synthesizers in parallel, or, equivalently, having an oracle that predicts which synthesizer is the best for a given instance and using it. The \ac{VBS} is able to solve more instances than any of the other synthesizers (92.7\% vs. the 79.4\% for \cubes). This shows two things: (1) not all synthesizers solve the same instances, and (2) it is advantageous to run multiple synthesizers in parallel if the user has the resources for it. Furthermore, if we consider a \ac{VBS} with only the top-performing synthesizers, \patsql and \cubes, the percentage of solved instances is 90.5\% (vs. 92.7\% with the four synthesizers), meaning that using two synthesizers in parallel results in 10\%+ extra instances solved compared to just using \cubes.

One interesting difference between these synthesizers is the minimum time in which they can return a solution for any of the instances, with \scythe and \patsql at around 0.3 seconds, while \squares and \cubes only solve the first instance at 2 to 3 seconds. The most likely explanation for this difference is the startup time for the programming languages used by the synthesizers. \patsql and \scythe both use Java, while \squares and \cubes use Python and also need to initialize the R execution environment.
\autoref{fig:seq-solved} also shows that both \scythe and \cubesseq
are able to solve more problem instances when we increase the time limit, while
\patsql and \squares seem to reach a plateau.

\autoref{table:results} shows the results for each benchmark set with a virtual
time limit of 10 seconds (top half) and a time limit of 10 minutes (bottom half). 
We can see that \cubesseq is able to solve more instances than \squares in all 
benchmarks, while solving more instances than \scythe in 3 out of 5 benchmark sets. 
When comparing with \patsql, the results shown in \autoref{fig:seq-solved}
are confirmed, since although \patsql solves more instances with a shorter time limit,
\cubesseq is able to solve more instances in one benchmark set (\texttt{spider}) with a larger time
limit.

\subsection{Parallel Performance using Simple Evaluation}
\label{subsec:parallel}

Considering the sequential version \cubesseq as our baseline, we now evaluate the 
performance of the parallel version using
divide-and-conquer (\cubesdcB).

\begin{table}[t]
\caption{Overall results for 10 seconds and 10 minutes, for all configurations tested, grouped by benchmark. The best tool for each time-limit/benchmark pair is highlighted in \textbf{bold}.} \label{table:results}
\centering
\vspace{4em}
\begin{tabular}{lccccccr}
 Run & \begin{rotate}{45}\texttt{kaggle}\end{rotate} & \begin{rotate}{45}\texttt{recent-posts}\end{rotate} & \begin{rotate}{45}\texttt{top-rated-posts}\end{rotate} & \begin{rotate}{45}\texttt{spider}\end{rotate} & \begin{rotate}{45}\texttt{textbook}\end{rotate} & All & \begin{tabular}[b]{@{}c@{}}Median\\Speedup\end{tabular} \\ 
  \midrule
  \multicolumn{8}{c}{10 seconds}\\
\textsc{Squares} & 21.2\% & 3.9\% & 5.3\% & 24.7\% & 28.6\% & 24.1\% \\ 
  \textsc{Scythe} & 0.0\% & \textbf{49.0\%} & \textbf{66.7\%} & 22.5\% & 28.6\% & 23.4\% \\ 
  \textsc{PatSQL} & \textbf{57.6\%} & 41.2\% & 64.9\% & 72.5\% & \textbf{62.9\%} & 71.7\% \\ 
  \textsc{Cubes-Seq} & 15.2\% & 11.8\% & 33.3\% & 51.5\% & 34.3\% & 50.3\% \\ 
  \textsc{Cubes-DC4} & 24.2\% & 11.8\% & 59.6\% & 70.0\% & 48.6\% & 68.5\% \\ 
  \textsc{Cubes-DC8} & 27.3\% & 15.7\% & 63.2\% & 73.2\% & 54.3\% & 71.8\% \\ 
  \textsc{Cubes-DC16} & 24.2\% & 19.6\% & 63.2\% & \textbf{75.4\%} & 51.4\% & \textbf{73.8\%} \\ 
   \midrule
   \multicolumn{8}{c}{10 minutes}\\
\textsc{Squares} & 21.2\% & 7.8\% & 22.8\% & 31.0\% & 40.0\% & 30.6\% \\ 
  \textsc{Scythe} & 3.0\% & \textbf{66.7\%} & \textbf{80.7\%} & 49.1\% & 54.3\% & 49.5\% \\ 
  \textsc{PatSQL} & \textbf{63.6\%} & 45.1\% & 66.7\% & 75.8\% & 68.6\% & 75.1\% \\ 
  \textsc{Cubes-Seq} & 39.4\% & 25.5\% & 66.7\% & 80.9\% & 57.1\% & 79.4\% & \textcolor{gray}{(\(1\,\times\))}\negphantom{)} \\ 
  \textsc{Cubes-DC4} & 45.5\% & 31.4\% & 73.7\% & 88.4\% & 71.4\% & 86.9\% & \(8.4\,\times\) \\ 
  \textsc{Cubes-DC8} & 54.5\% & 39.2\% & 73.7\% & 89.6\% & 68.6\% & 88.2\% & \(12.8\,\times\) \\ 
  \textsc{Cubes-DC16} & 51.5\% & 39.2\% & 75.4\% & \textbf{90.4\%} & \textbf{77.1\%} & \textbf{89.0\%} & \(15.5\,\times\) \\ 
   \bottomrule
\end{tabular}
\end{table}

\subsubsection{Divide-and-Conquer}

\autoref{table:results} shows the results for the divide-and-conquer strategy
\cubesdcB with 4, 8 and 16 processes. Notice that divide-and-conquer tools 
improve upon the sequential version, from 79.4\% up to 89.0\% when using 16
processes. Moreover, within a limit of 10 seconds, the parallel versions are
able to solve 68.5\%, 71.8\% and 73.8\% of the instances, when using respectively
4, 8 and 16 processes. This contrasts with the sequential version only solving 50.3\% when a 10 second limit is imposed. This shows that there is
a significant speedup when using the divide-and-conquer strategy, especially when using shorter time limits.

Formally, the speedup of method $A$ in relation to method $B$ is defined as the 
time needed to execute method $B$ divided by the time 
needed to execute method $A$, and is a measure of how fast an implementation 
is compared to another.
The last column of \autoref{table:results} shows the speedup obtained by each parallel version
of \cubes in relation to the sequential version \cubesseq for instances where
\cubesseq needed 1 minute (or more) to solve. We focus this analysis
on the harder instances for the sequential tool, since higher speedups in
these instances have a higher impact on the end user's experience.

\begin{figure}[t]
    \centering
    \includegraphics{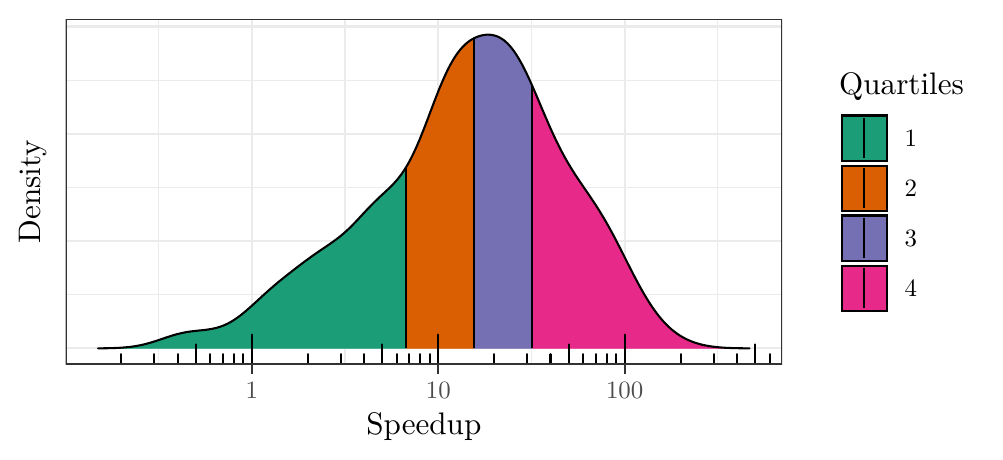}
    \caption{Instance speedup distribution for \cubesdc{16}.} \label{fig:speedup}
\end{figure}

We can see that most configurations have a median speedup greater than the number of 
processes used. This is called a super-linear speedup and occurs because programs 
are enumerated in a different order when using our parallel versions. 
\autoref{fig:speedup} shows the full speedup distribution for \cubesdc{16} along with 
the distribution quartiles. 
We can see that more than 50\% of instances have a speedup greater than 10 
when using 16 processes, while more than 25\% of instances have a speedup greater 
than 30.

\subsubsection{Ablation Study}
In order to evaluate the impact of the different techniques introduced in Sections~\ref{sec:sequential} and~\ref{sec:parallel}, we perform an ablation study, which is presented in \autoref{table:ablation}.
In this table, ``No Invalid Program Pruning'' represents an execution where the invalid program pruning introduced in \autoref{sec:cubes-deducing} was disabled, keeping all other features enabled.
Likewise, ``No Split Program Space'' shows an execution where the program space splitting introduced in Section~\ref{sec:split} was disabled.
Finally, ``Random Cube Generation'' corresponds to fully random cube generation
instead of using the dynamic cube generation -- this implies disabling the program space splitting as well.

Looking at the results, we can see that all three of these features have a significant impact when looking at the number of instances solved in the first 10 seconds, with an average of 66.9\% instances solved versus 73.8\% when all features are enabled. Meanwhile, looking at the full 10 minutes, only the invalid program pruning shows a significant impact. This shows that while all 3 features allow cubes to produce answers faster, only the invalid program pruning allows cubes to produce more answers under a 10-minute limit. The reason why this happens is that while the invalid program pruning technique prunes areas of the program space, effectively discarding programs that will not lead to a solution, the other two techniques only try to guide the search towards areas of the program space where a solution is likely to be found. With larger time limits, these techniques become less meaningful.
Furthermore, \autoref{table:ablation} also shows that disabling the program space splitting positively affects the number of instances solved in 10 minutes. This might be due to the percentage of available processes assigned to each part of the program space, as described in Section~\ref{sec:split}. One possible improvement to this technique would be to use a dynamic approach for assigning threads to each DSL splitting section.

Finally, the random cube generator results are fairly similar to the results when the program splitting is disabled. This would indicate that the dynamic cube generator is not very significant for \cubes' performance. However, program space splitting and the dynamic cube generator must be enabled to obtain good performance at lower time limits.

\begin{table}[t]
\caption{Ablation study results for 10 seconds and 10 minutes grouped by benchmark. The best tool for each time-limit/benchmark pair is highlighted in \textbf{bold}.} \label{table:ablation}
\centering
\vspace{4em}
\setlength{\tabcolsep}{3pt}
\begin{tabular}{lcccccc}
 Run & \begin{rotate}{45}\texttt{kaggle}\end{rotate} & \begin{rotate}{45}\texttt{recent-posts}\end{rotate} & \begin{rotate}{45}\texttt{top-rated-posts}\end{rotate} & \begin{rotate}{45}\texttt{spider}\end{rotate} & \begin{rotate}{45}\texttt{textbook}\end{rotate} & All \\ 
  \midrule
  \multicolumn{7}{c}{10 seconds}\\
\textsc{Cubes-DC16} (Default) & 24.2\% & 19.6\% & 63.2\% & \textbf{75.4\%} & \textbf{51.4\%} & \textbf{73.8\%} \\ 
  No Invalid Program Pruning & 31.6\% & 21.6\% & 64.9\% & 69.1\% & 42.9\% & 68.0\% \\ 
  No Split Program Space & 44.4\% & \textbf{25.5\%} & 63.2\% & 67.6\% & 48.6\% & 66.7\% \\ 
  Random Cube Generation & \textbf{46.7\%} & 21.6\% & \textbf{66.7\%} & 66.9\% & 48.6\% & 66.1\% \\ 
   \midrule
   \multicolumn{7}{c}{10 minutes}\\
\textsc{Cubes-DC16} (Default) & 51.5\% & 39.2\% & 75.4\% & 90.4\% & \textbf{77.1\%} & 89.0\% \\ 
  No Invalid Program Pruning & 57.9\% & 33.3\% & 75.4\% & 86.3\% & 60.0\% & 85.1\% \\ 
  No Split Program Space & 55.6\% & \textbf{43.1\%} & \textbf{78.9\%} & \textbf{90.8\%} & 71.4\% & \textbf{89.7\%} \\ 
  Random Cube Generation & \textbf{60.0\%} & \textbf{43.1\%} & \textbf{78.9\%} & 90.4\% & 71.4\% & 89.3\% \\ 
   \bottomrule
\end{tabular}
\end{table}

\subsection{Effect of Input Structure on Performance} \label{subsec:io-structure-results}

In this section, we analyze the impact of several features from the input-output example on the performance of \cubes and the other synthesizers. In particular, we explore this issue considering the metrics of the input-output examples presented in Section~\ref{sec:data}.

\begin{figure}
    \centering
    \includegraphics{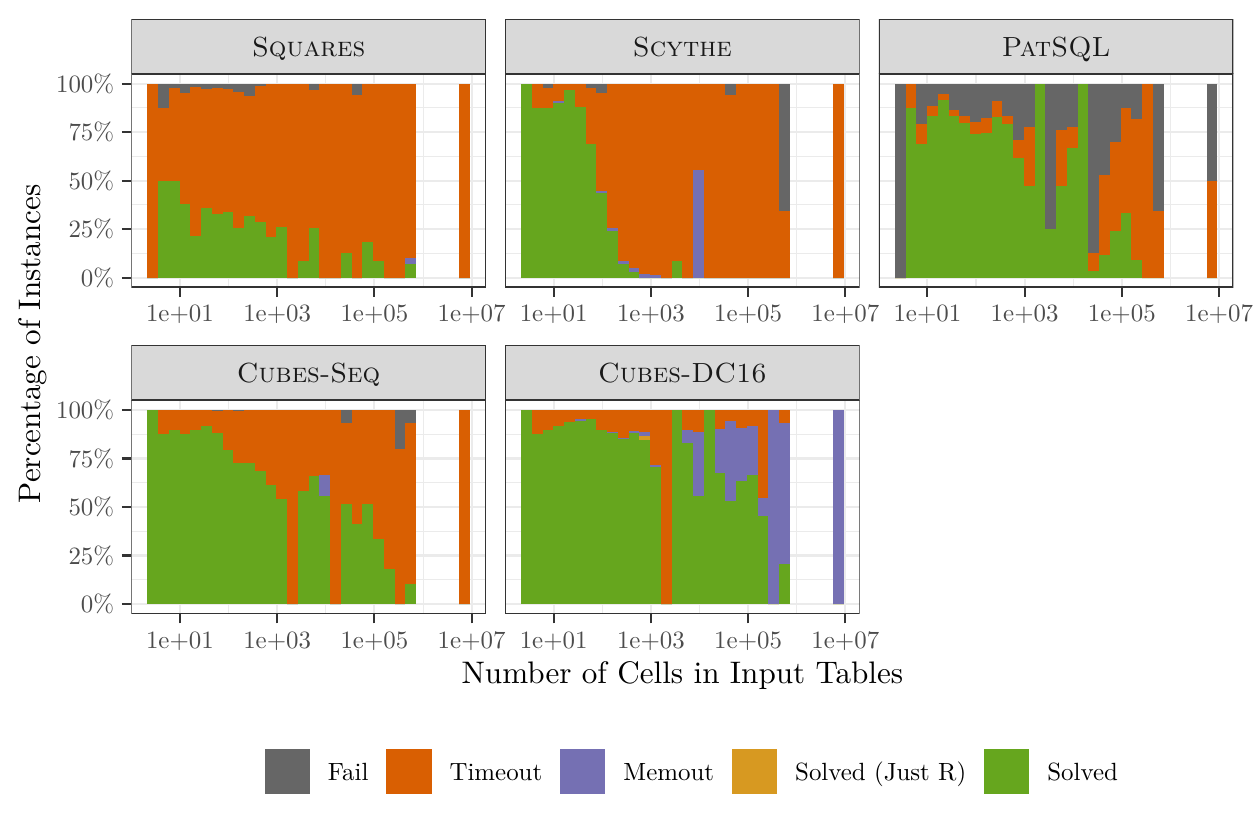}
    \caption{Percentage of instances solved as a function of the number of cells in the input tables, using simple evaluation.}
    \label{fig:solved_cells}
\end{figure}

\begin{figure}
    \centering
    \includegraphics{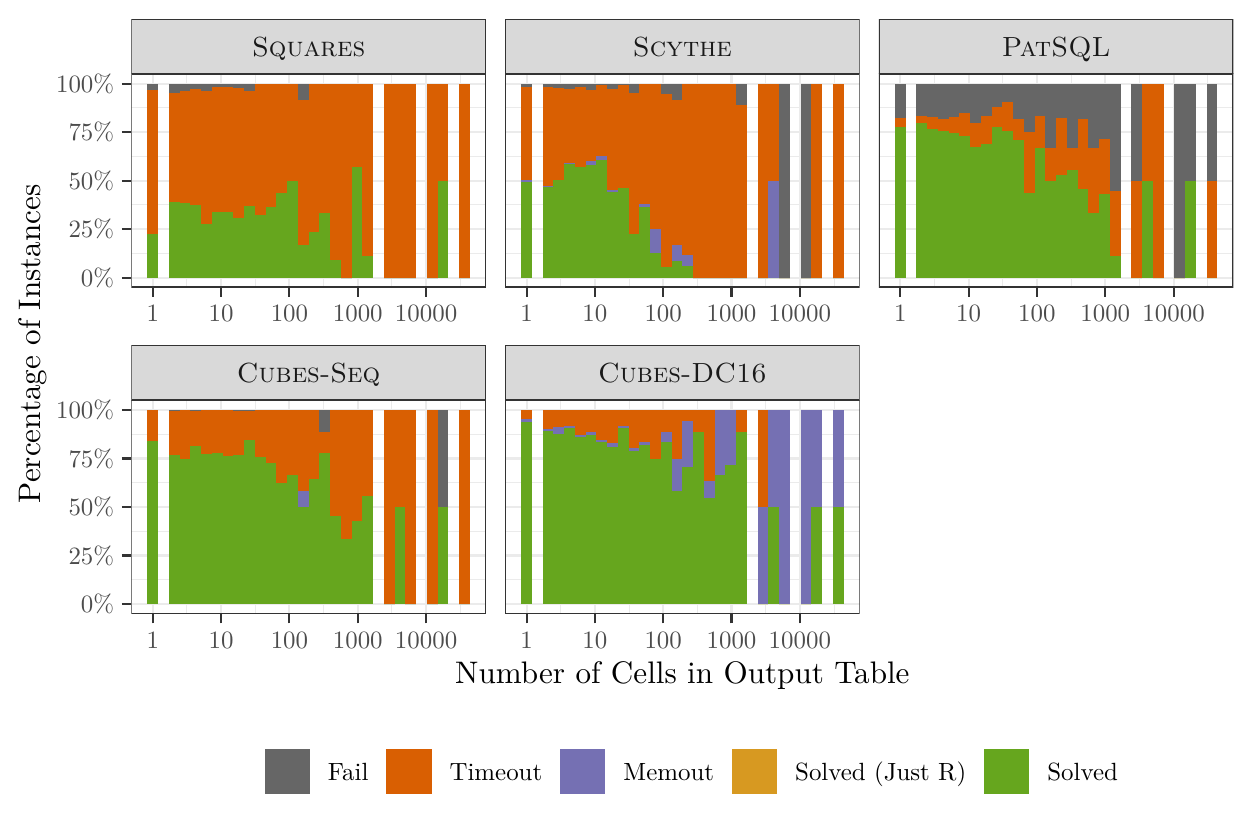}
    \caption{Percentage of instances solved as a function of the number of cells in the output table, using simple evaluation.}
    \label{fig:solved_output_cells}
\end{figure}

In \autoref{fig:solved_cells}, we explore the relationship between the number of cells in the input tables and the capability of the different synthesizers to find a solution for those instances. Instances marked as ``Fail'' mean that the corresponding synthesizer terminated with a non-zero exit code without providing a solution. Generally, all synthesizers solve a smaller percentage of instances with large databases than small ones. However, this is most pronounced in \scythe, which cannot solve any instance with more than 4000 cells in the input database. 
\cubesdc{16} is the synthesizer able to solve more instances with large input databases. However, there is also a great number of memouts for this type of instances. This is due to \cubesdc{16} having the same total available memory as the other synthesizers, despite using 16 processes.

In \autoref{fig:solved_output_cells}, we present a similar plot for the number of cells in the output table. Once again, synthesizers seem to have trouble finding solutions for instances where the output table is very large. Even so, this effect is not as pronounced as for the size of the input database. This is especially true for \cubesdc{16}, which seems to be the least affected by the output size. Once again, the most affected tool is \scythe, which cannot solve any instance with more than 200 cells in the output table.

\begin{figure}
    \centering
    \includegraphics{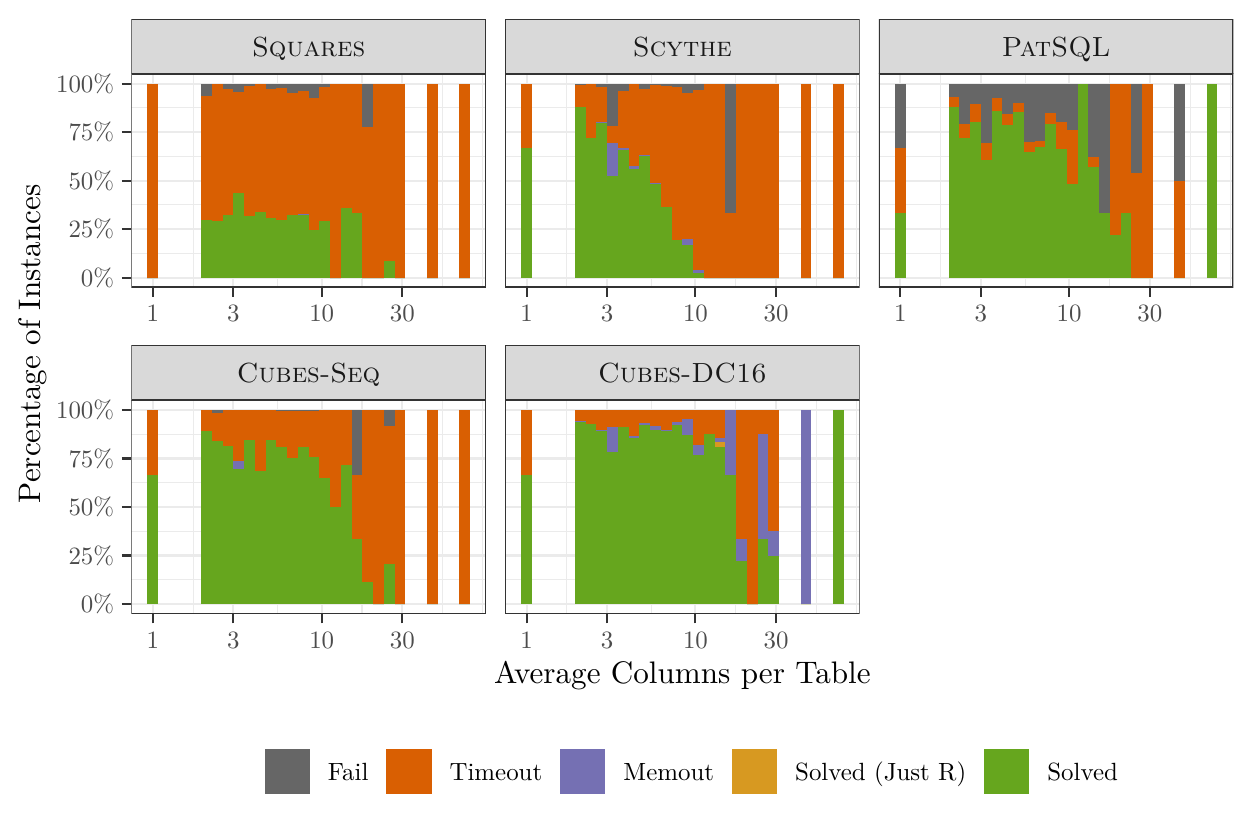}
    \caption{Percentage of instances solved as a function of the average number of columns per input table, using simple evaluation.}
    \label{fig:solved_cols_per_table}
\end{figure}

\autoref{fig:solved_cols_per_table} shows the effect of the average number of columns per input table on the synthesizer performance. \scythe and \cubesseq solve fewer instances with a higher number of columns, with the effect being more pronounced in \scythe. Meanwhile, \patsql and \cubesdc{16} are not as affected, being the only synthesizers capable of solving an instance with an average of 70 columns per input table.

\subsection{Results using Fuzzing-based Evaluation}
\label{subsec:accuracy}

\begin{figure*}[t]
    \centering
    \includegraphics{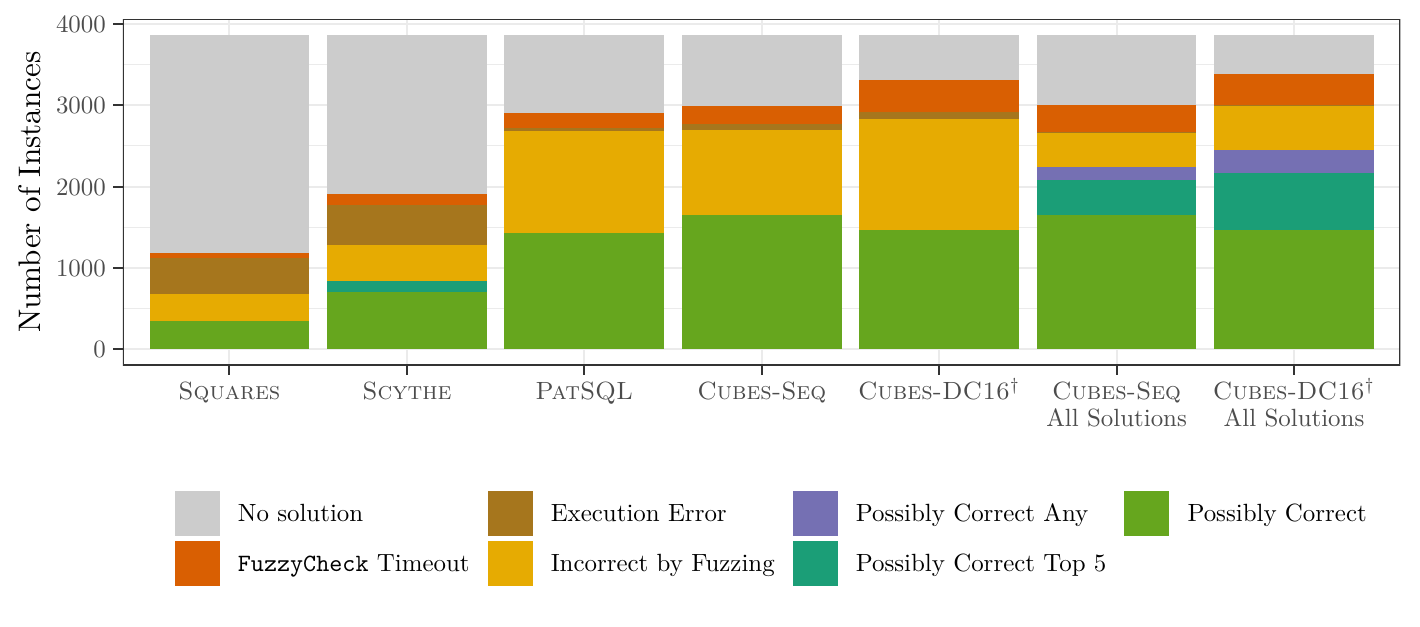}
    \caption{Results of the fuzzy-based evaluation for each synthesizer. Results for the runs marked with \textdagger{} differ due to non-determinism effects in \cubesdc. These effects are further discussed in \autoref{sec:non-determinism}.} \label{fig:fuzzy}
\end{figure*}

\begin{figure*}[t]
    \centering
    \includegraphics{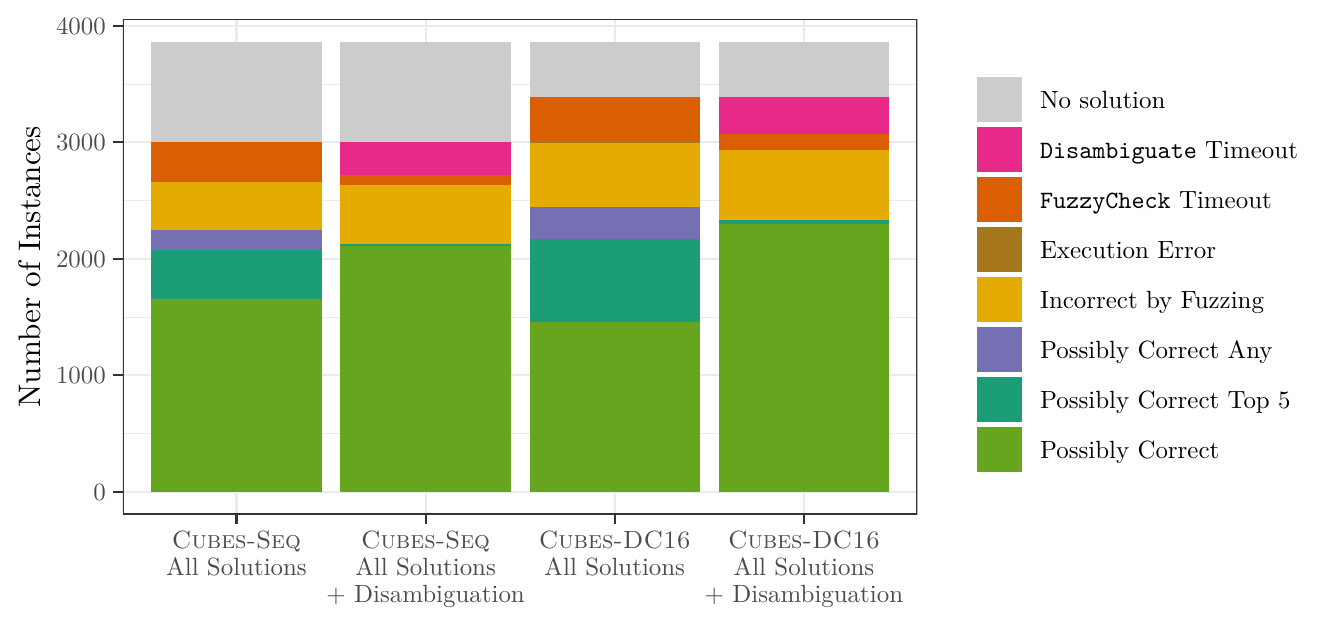}
    \caption{Comparison of the fuzzy-based evaluation results before and after disambiguation.} \label{fig:disambigation-results}
\end{figure*}

\begin{table}
\small
\caption{Comparison of the fuzzy-based evaluation with the simple evaluation.}
\label{tab:fuzzy}
\begin{minipage}{\columnwidth}
\center
\begin{tabular}{@{}lrrrrr@{}}
\toprule
          & \scythe & \squares & \patsql & \begin{tabular}[t]{@{}c@{}}\cubesseq\\All Solutions\end{tabular} & \begin{tabular}[t]{@{}c@{}}\cubesdc{16}\\All Solutions\end{tabular} \\ \midrule
Solved (simple eval.)    & 49.5\%   & 30.6\%    & 75.1\%   & 79.5\% & 90.2\%  \\[.2ex] \hdashline
Possibly Correct\footnote{Includes instances in Possibly Correct Top 5 and Possibly Correct Any.} \rule{0pt}{2.4ex}   & 21.6\%   & 9.2\%     & 37.1\%   & 58.0\% & 63.3\%  \\
\hspace{1em}{\small\color{gray}\emph{as \% of Solved instances}}   & {\small\color{gray}43.6\%}   & {\small\color{gray}30.0\%}     & {\small\color{gray}49.4\%}   & {\small\color{gray}73.0\%} & {\small\color{gray}70.2\%}  \\

Incorrect by Fuzzing & 11.6\%   & 8.4\%     & 32.3\%   & 10.7\% & 14.1\%  \\
\hspace{1em}{\small\color{gray}\emph{as \% of Solved instances}}   & {\small\color{gray}23.4\%}   & {\small\color{gray}27.5\%}     & {\small\color{gray}43.0\%}   & {\small\color{gray}13.5\%} & {\small\color{gray}15.6\%}  \\

Inconclusive & 16.2\%   & 13.1\%     & 5.7\%   & 8.9\% & 10.2\%  \\
\hspace{1em}{\small\color{gray}\emph{as \% of Solved instances}}   & {\small\color{gray}32.7\%}   & {\small\color{gray}42.8\%}     & {\small\color{gray}7.6\%}   & {\small\color{gray}11.2\%} & {\small\color{gray}11.3\%}  \\\bottomrule
\end{tabular}
\end{minipage}
\end{table}

In this section we analyze the number of instances solved by \cubes when using the more thorough fuzzy-based evaluation, as well as comparing it with other program synthesis tools. Furthermore, we also evaluate the program disambiguator introduced in \autoref{sec:disambiguation}.

\autoref{fig:fuzzy} shows the results when using the fuzzy-based evaluation method instead of the simple evaluation. For this evaluation, we used 16 fuzzing rounds (\(R = 16\)). The ``\texttt{FuzzyCheck} Timeout'' label in the plot represents instances for which the fuzzing evaluation timed out and not a timeout of the synthesizer used. We used a time limit of 60 seconds per fuzzing round (\(16\times 60\mathrm{s} = 960\mathrm{s}\)). Furthermore, some of the synthesized queries failed to execute (labelled as ``Execution Error''). This happens for two reasons: (1) some synthesized queries are incompatible with the SQLite dialect, and (2) some of the synthesized queries contain syntax problems.

We label instances for which we could not find a distinguishing input from the ground truth as ``Possibly Correct'', while instances for which we did find such input are labelled as ``Incorrect by Fuzzing''. 
Furthermore, for synthesizers that return multiple solutions, ``Possibly Correct Top 5'' means that there was a query in the top-5 returned queries for which we did not find a distinguishing input from the ground truth. Similarly, ``Possibly Correct Any'' means that the synthesizer returned a query for which we could not distinguish it from the ground truth.

Previous tools all suffer from fairly low accuracy rates, staying under 45\%, as do \cubesseq and \cubesdc{16} if we only consider the first solution returned. However, if we consider all solutions returned under 10 minutes, then \cubes generates a correct (using fuzzy-based evaluation) solution on around 63\% of the instances, as shown in \autoref{tab:fuzzy}.

In order to be able to give that correct solution to the user, as opposed to giving them all the solutions generated, we developed a query disambiguator. \autoref{fig:disambigation-results} shows the results of using that disambiguator on \cubesseq and \cubesdc{16}. We can see that the disambiguator can almost always identify the correct query if such a query exists in the set of queries synthesized. Note that small differences in the exact number of queries deemed correct using the fuzzy-based evaluation may be due to different fuzzed inputs being generated.

It is also worth noting that a very small number of instances are labeled as ``Possibly Correct Top 5''.
As explained in Section~\ref{sec:disambiguation}, \cubes returns the earliest synthesized query when we reach a set of queries that we cannot distinguish from one another. This means that, for those instances, a correct query was in the final set of queries selected by the disambiguation, but it was not the first one generated by \cubes. This happens because while the accuracy test has access to the ground truth and can thus generate better-fuzzed inputs, the disambiguator is limited to using values from the queries it is trying to disambiguate. Even so, the fact that this only occurs in a very small number of queries indicates that the approach is valid and seems to be able to both correctly disambiguate most queries and catch the cases where the disambiguation fails.

We show that if we only consider the first solution, \cubes' performance is similar to other existing tools. The main improvement comes from (1) synthesizing many possible queries for a given problem and (2) having a program disambiguator to choose the right query. This first point is directly influenced by our parallel approach to program synthesis, which allows us to synthesize more programs that satisfy the examples under the chosen time limit.

\begin{figure}[t]
    \centering
    \includegraphics{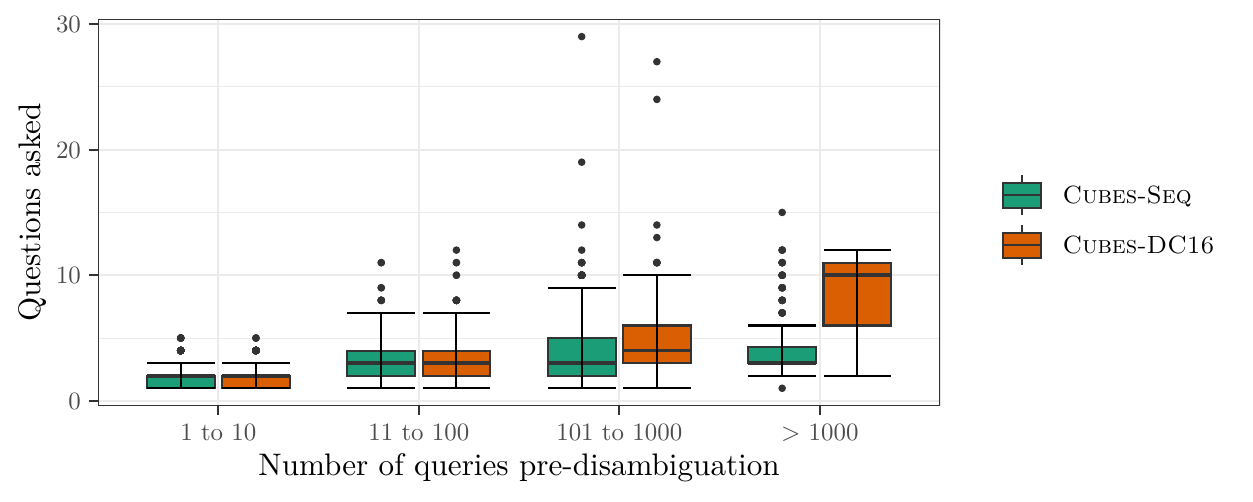}
    \caption{Number of questions that need to be asked to the user in order to perform disambiguation, as a function of the number of queries synthesized. Points represent instances that are potential outliers, using \(1.5\cdot\mathrm{IQR}\).} \label{fig:n_questions}
\end{figure}

Finally, we analyze how many questions are asked to the user to disambiguate the queries produced by \cubes. \autoref{fig:n_questions} shows this data as a function of the number of queries synthesized. Consider the first bar of the second group, relating to instances where \cubesseq generated 11 to 100 queries. The plot shows that to disambiguate those queries, we need at least 1 question, at most 11 questions, and on average 3 questions.

For \cubesseq the average number of questions needed to disambiguate up to 1000 queries is 2.31, while for \cubesdc{16} it is 2.69. As stated in Section~\ref{sec:disambiguation}, our goal with the disambiguation strategy is to discard half the queries with each question asked. Thus, we would expect that the number of questions needed to disambiguate a given set of queries scales logarithmically with the size of that set. \autoref{fig:n_questions} shows that this behavior is, in fact, observed in practice.

\subsection{Non-determinism}
\label{sec:non-determinism}

The parallel version of \cubes is non-deterministic. This means that,  
if run several times, it might not always produce the same queries
or solve the same benchmarks within the time limit. 
In order to evaluate this non-deterministic behavior, we randomly chose a subset 
consisting of 500 instances
and executed \cubesdc{16} 10 times on each of 
them, in order to analyze the variance in \cubes' behavior. 

\begin{figure}[t]
    \centering
    \includegraphics{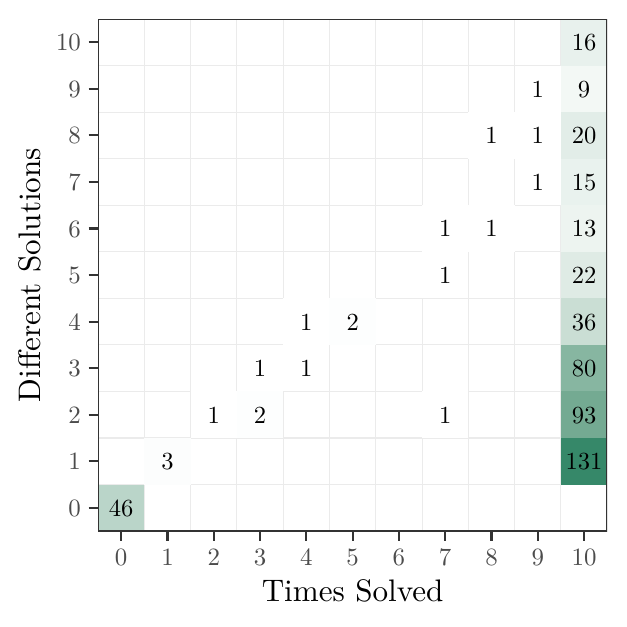}
    \caption{Non-deterministic behavior of \cubesdc{16}. A random selection of 500 instances was run 10 times each.} \label{fig:non-determ}
\end{figure}

\autoref{fig:non-determ} shows a summary of the results. In this 
matrix, the $x$ axis shows how many times (out of 10) a solution was found for a given 
instance, while the $y$ axis shows how many different solutions were found for a given 
instance. For example, the number 80 in position (10,3) of the matrix means that 
there were 80 instances out of 500 that were solved 10/10 times, and 
that out of those 10 times there were 3 different solutions overall.

We can see that only 19 (4\%) out of 500 instances are only solved in some of
the 10 runs. Hence, the non-determinism of \cubesdc{16} has a very low impact 
on the ability to provide an answer to the user.
Furthermore, out of the 435 problem instances that were always solved, 131 (26.2\%) always produce the same exact query, and 93 (18.6\%) produce two different queries in 10 runs. Therefore, a non-deterministic behavior is observed, but in only 16 instances (3.2\%) do we get a different query each time the tool is executed. These rare cases are due to a clearly under-specification of the user intent (i.e., the input-output example can be easily satisfied by many different queries). 

\section{Discussion}
\label{sec:limitations}

Here we discuss the main threats to validity of this work and some challenges that were raised during the experimental evaluation.

\paragraph{Benchmarks} Our evaluation uses a large set of benchmarks from different domains. However, they may not be representative of tasks commonly performed by users or may have a bias towards a specific synthesis tool. To mitigate this, we included benchmarks from several previous synthesis tools and also extended a large dataset from query synthesis using NLP to use examples instead. In the end, we have around 4000 instances but they are dominated by the \texttt{spider} dataset~\cite{spider}. Nevertheless, since this dataset has been extensively used in other domains and was not created by us, we believe that it is more general and less prone to bias.

\paragraph{Parallelism} The divide-and-conquer approach already shows scalability for hard instances when using 4 and 8 processes in a multicore architecture with super-linear speedups. However, when increasing the number of processes to 16 the gains are reduced. When the number of processes increases, there is an increase of contention for memory accesses that can slow down the performance of each process. To address this issue, it would be interesting to evaluate \cubes in a distributed setting. Note that the overhead of going from multicore to distributed should be small since the inter-process communication is already done using message-passing techniques, and no shared memory is used. 
Exchanging information between processes is another source of improvement that would be worth exploring in future work.

\paragraph{Cube generation} One way to further improve the divide-and-conquer approach is to consider other cube generation strategies. For instance, we could learn from data and use machine learning techniques such as pre-trained bigram scores or using neural networks to predict the most likely cubes. We could also explore other techniques similar to the ones used in SAT solvers, such as restarting the search after \(n\) programs/cubes have been attempted. 

\paragraph{Determinism} \cubesdc{} is currently non-deterministic, but if used in an industrial setting, it may be important to have a deterministic version. This could be achieved at the cost of performance by (i) updating the bigram scores in batches and in a deterministic way, (ii) solving cubes in batches so that processes stay synchronized --- this would require that cubes be of approximately the same difficulty to reduce stalls, and (iii) find a deterministic way to assign generated cubes to the available processes. When using \cubes with disambiguation, we enumerate all possible solutions within a given time limit. In this setting, non-determinism is not a major issue. Even though the order of enumeration may differ, we will likely enumerate the same set of solutions in different synthesizer runs.

\paragraph{Fuzzy-based Evaluation} Even though query synthesis tools are becoming more efficient and can find a query that satisfies the input-output example given by the user, they may not find the query that the user intended. To the best of our knowledge, this is the first study where fuzzing was used to evaluate if the query returned by the synthesizer matches the user's intent.
Even though fuzzing is not a precise measurement of correctness since it may return that some queries are equivalent when they may not be, it is an upper bound on the accuracy of these tools. With the continuous improvement of SQL equivalence tools
~\cite{DBLP:conf/cidr/ChuWWC17, DBLP:journals/pvldb/ZhouANHX19, DBLP:journals/pvldb/ChuMRCS18}, it may be possible to have an exact accuracy measurement in the future. 
However, even with the current results, we already observe that all synthesis tools return many answers that do not match the desired behavior. 

\paragraph{Disambiguation} Interacting with the user to perform query disambiguation is essential to increase the accuracy of SQL synthesizers based on examples. However, the questions that we asked the user may be too hard to answer, or the user may answer them incorrectly. To mitigate the difficulty of the questions, we only ask yes or no questions and present examples based on fuzzing that are often similar to the initial example provided by the user. With this approach, we hope that the user can quickly answer these questions. We currently automate the disambiguation procedure and use the ground truth to answer the questions, but a user study could be done in the future to confirm our hypothesis that the questions are easy for users to answer. In this work, we assume that the user never answers the questions incorrectly. However, considering this scenario could open new research directions and is in line with recent work on program synthesis with noisy data~\cite{DBLP:conf/sigsoft/HandaR20} where the examples may be incorrect.

\section{Related Work}
\label{sec:related}

\paragraph{SQL Synthesis.}
In recent years, several tools for query synthesis have been proposed using input-output 
examples to specify user intent~\cite{sql-output,sqlsynthesizer,morpheus,scythe,trinity,orvalhoVLDB20}.
Solving approaches vary from using decision trees with fixed 
templates~\cite{sql-output,sqlsynthesizer}, to abstract representations of queries 
that can potentially satisfy the input-output examples~\cite{scythe}.
Another approach is to use SMT-based representations of the search 
space~\cite{morpheus,orvalhoCP19} such that each solution to the SMT formula
represents a possible candidate query to be verified.
The \cubes framework proposed in this paper is also based on SMT-based representations but it extends prior work in several dimensions: (i) extends the language in the programs to be synthesized,
(ii) proposes pruning techniques that can be directly encoded into SMT,
and (iii) it is the first parallel tool for query synthesis.

In this paper, we compare \cubes with three other \ac{SQL} Synthesis tools that use input-output examples: \scythe~\cite{scythe}, \squares~\cite{orvalhoVLDB20} and \patsql~\cite{patsql}. \scythe and \patsql use sketch-based enumeration, where first a skeleton program with missing parts is generated and then, if the skeleton satisfies a preliminary evaluation, the synthesizer tries to complete the sketch to obtain a full program. \squares, on the other hand, uses \ac{SMT}-based enumeration where full programs are obtained by iterating the possible solutions of an \ac{SMT} formula. Both \scythe and \squares have limited \acp{DSL} and thus are not as well suited for complex tasks. Furthermore, \scythe's ability to solve a given instance is severely limited by the size of its input tables. Although \patsql has a comparatively more expressive \ac{DSL}, it is still not able to outperform \cubes.

Another approach to specify user intent is the use of natural language~\cite{sqlizer,ratsql}.
However, these approaches often need a large training data set from the query's domain.
Recently, several techniques have been proposed that try to better generalize to 
cross-domain 
data~\cite{DBLP:journals/corr/abs-2009-13845,DBLP:journals/corr/abs-2012-10309}. 
Although many improvements have been attained on finding the structure of the
query through effective semantic table parsing, defining the details (e.g. specific filter 
conditions) is usually hard, in particular in more complex queries. The use of natural language for query synthesis is complementary to our approach and a combination of both strategies could improve the accuracy of program synthesizers at the cost of more input from the user, namely examples and a natural language description of the task.

\paragraph{Program Disambiguation.}
Current synthesizers focus mostly on generating programs that satisfy the user's specification.
However, in many situations, the produced program does not satisfy the true user intent \cite{DBLP:conf/uist/MayerSGLMPSZG15, DBLP:conf/icse/ShriverES17}. 
Previous work has shown that this shortcoming can be solved without recurring to complete specifications by introducing a program disambiguator. This component is responsible for interacting with the user and choosing between several possible solutions. \textcite{DBLP:conf/uist/MayerSGLMPSZG15} describe two types of user interaction for program disambiguation: in the first approach, users select the correct program among a set of returned solutions which are presented in a way that allows easy navigation. The second approach is described as \emph{conversational clarification}, where the system iteratively asks questions to the user, further refining the original specification until just one candidate program is left \cite{DBLP:conf/tacas/FerreiraTVLM21, DBLP:conf/kbse/RamosPLMM20, DBLP:journals/pvldb/LiCM15, DBLP:conf/sigmod/WangCB17, DBLP:conf/pldi/JiLXZH20, DBLP:conf/iui/NaritaMLI21}. In \cubes, we use conversational clarification to improve the confidence in produced solutions while still keeping the complexity for the user low.

\paragraph{Parallel Solving.}
Solving logic formulas in parallel has been the subject of extensive research
work~\cite{DBLP:books/sp/HS2018,DBLP:journals/tplp/GentMNMPMU18,DBLP:conf/sat/AignerBKNP13,DBLP:conf/sat/BalyoSS15}, both using 
memory-shared~\cite{DBLP:journals/informs/ShinanoHVW18} and 
distributed approaches~\cite{DBLP:journals/amcs/NgokoCT19}. One of the techniques used to explore the search space is called divide-and-conquer~\cite{DBLP:books/sp/18/HeuleKB18}.
In this approach, the search space is split into
disjoint areas such that there is no intersection between the areas explored
by each process. In this case, work stealing techniques~\cite{Schubert05} are commonly
used to avoid starvation, since the search space can be unevenly split among
the processes.
Although we adapt techniques from parallel automated reasoning, the parallelization
in the \cubes framework is not done at solving logic formulas, but in a more 
abstract level. In our case, logic formulas continue to be solved sequentially.
Moreover, starvation is avoided by producing additional work, i.e. increasing the
number of operations from the DSL in the programs to be enumerated.

\section{Conclusions}
\label{sec:conc}

This work introduces \cubes, a new enumeration-based framework for query synthesis from examples. A new robust synthesizer is proposed that extends the range of produced SQL queries. Moreover, it also includes new pruning techniques that allow the cutting of invalid programs from the enumeration process, resulting in an effective new tool for SQL synthesis in a broad range of benchmark sets.
Additionally, \cubes also takes advantage of the current multicore processor architectures, providing the first parallel query synthesizer from examples using  a divide-and-conquer approach. The splitting of the program space is done by providing different sequences of operations to each thread, as well as performing DSL splitting among threads.

An in-depth experimental evaluation is also carried out, comparing \cubes with other state-of-the-art query synthesizers in a wide variety of benchmark sets.
Experimental results show the effectiveness and robustness of \cubes, being able to successfully synthesize SQL queries for a larger range of problem instances than other tools. 
Moreover, the parallel versions of \cubes have super-linear speedups for many hard instances and, when using 16 processes, provide a median speedup of $15\times$ over the sequential version of the tool.

Finally, an accuracy analysis of the produced queries is also performed using 
fuzzing techniques. Results show that the queries produced by current synthesizers  often differ from the user intent. Therefore, an interactive procedure with the user is proposed to disambiguate among all queries found by \cubes that satisfy the original input-output example. After the disambiguation procedure, the accuracy of \cubes in providing the user intent query is greatly increased from around 40\% to 60\%.

\begin{acks}
This work was partially supported under National Science Foundation Grant No. CCF-1762363, by OutSystems and by Portuguese national funds through FCT, under projects UIDB/50021/2020, PTDC\-/CCI-COM/\-31198/\-2017, DSAIPA\-/AI/0044/2018, and project ANI 045917 funded by FEDER and FCT. Support was also provided by the Fundação para a Ciência e a Tecnologia (Portuguese Foundation for Science and Technology) through the Carnegie Mellon Portugal Program under Grant PRT/BD/152086/2021.
\end{acks}

\bibliographystyle{ACM-Reference-Format}
\bibliography{bibliography}

\end{document}